\documentclass[english,aps,showpacs,preprintnumbers,amsmath,amssymb,nofootinbib,12pt]{revtex4}

\usepackage{graphicx}
\usepackage{color}

\newcommand{\STr}{\mathrm{STr}}

\newcommand{\E}{\mathrm{e}}
\newcommand{\I}{\mathrm{i}}
\newcommand{\be}{\begin{eqnarray}}
\newcommand{\ee}{\end{eqnarray}}

\newcommand{\nn}{\nonumber }

\newcommand{\fderL}[1]{\frac{\overrightarrow{\delta}}{\delta #1}}
\newcommand{\fderR}[1]{\frac{\overleftarrow{\delta}}{\delta #1}}

\usepackage{babel}
\makeatother

\begin{document}

\title{The QCD Phase Boundary from Quark-Gluon Dynamics}
\date{\today}                                           
\author{Jens~Braun}
\affiliation{TRIUMF, 4004 Wesbrook Mall, Vancouver, BC V6T 2A3, Canada}

\begin{abstract}
We study one-flavor QCD at finite temperature and chemical potential using the functional
renormalization group. We discuss the chiral phase transition in QCD and its order with its underlying mechanism
in terms of quarks and gluons and analyze the dependence of the phase transition temperature on small quark 
chemical potentials. Our result for the curvature of the phase boundary at small quark chemical potential
relies on only a single input parameter, the value of the strong coupling at the $Z$ mass scale.
\end{abstract}


\maketitle

\section{Introduction}
The phase boundary of Quantum Chromodynamics (QCD) is currently a very active frontier both theoretically and
experimentally. For fixed small quark chemical potentials, the ground-state of QCD changes with increasing temperature from a hadronic
phase with dynamically broken chiral symmetry to a deconfined quark-gluon plasma phase with an effectively 
restored chiral symmetry. Even though the phase transition temperatures are not directly observable in heavy-ion collision experiments at BNL
and CERN, a lower bound can be extracted from the experimental data~\cite{BraunMunzinger:2003zz}. These so-called chemical 
freeze-out temperatures can then be compared to theoretical predictions for the chiral and deconfinement phase-transition temperature.
Since QCD is a strongly-interacting theory and long-range fluctuations need to be captured in order to study phase transitions,
non-perturbative approaches are indispensable for a study of the QCD phase boundary.

On the theoretical side, various approaches are available for studies of the QCD phase boundary,  e. g. lattice QCD
simulations or functional methods. Each of these approaches comes with advantages
and disadvantages. Lattice QCD simulations are certainly the most powerful tool for a study of full QCD. However, 
the implementation of chiral fermions continues to be a non-trivial task. At finite 
chemical potential, the spectrum of the Dirac operator becomes complex, making direct lattice simulations even more difficult. In the past decade, 
however, several methods have been developed to circumvent the problems arising at finite chemical potential, such as studies of QCD 
at imaginary chemical potential~\cite{deForcrand:2002ci,D'Elia:2002gd,deForcrand:2003hx,deForcrand:2006pv}, 
Taylor expansions of the path integral or reweighting techniques~\cite{Fodor:2001au,Karsch:2003va,Fodor:2004nz,Fodor:2007pg,Gavai:2008zr},
see e.~g. Refs.~\cite{Schmidt:2006us,Philipsen:2008gf} for short overviews.

Functional approaches to QCD, such as mean-field studies, Dyson-Schwinger Equations or Renormalization Group approaches, do not 
have problems arising from a discretized action or a complex-valued spectrum of the Dirac operator. However, a study of full QCD is not possible
and a truncation of the QCD action functional is unavoidable. Therefore Lattice QCD simulations and continuum 
approaches should be considered as complementary approaches for studies of the QCD phase diagram.

Dynamical chiral symmetry breaking has been studied by applying effective low-energy models such 
as the Nambu-Jona-Lasinio (NJL) model~\cite{Nambu:1961tp}. The application of these models is built on the assumption 
that QCD falls into a certain universality class, namely $O(4)$. Whether this assumption is justified or not is currently under investigation 
by Lattice QCD simulations as well as functional RG methods~\cite{D'Elia:2005bv,Aoki:2006we,Braun:2007td,Klein:2007qh,Braun:2008sg}. 

Although NJL-type models already allow to study dynamical chiral symmetry breaking at finite temperature and quark chemical potential,
they do not contain gluonic degrees of freedom and they are not confining. Moreover, an ultraviolet (UV) cutoff has to be introduced in 
the theory. This makes the connection of these models to high-momentum scales and temperatures difficult. 
The dependence on the UV cutoff implies a parameter dependence of the model. The strategy for employing these models 
is usually as follows: First, one uses a set of parameters
and the UV cutoff to fit the values of low-energy observables at zero temperature and zero chemical potential, e. g. to the pion 
mass and to the pion decay constant. Second, one computes the phase boundary of QCD while keeping the parameters and the UV cutoff 
fixed. A shortcoming of these models is apparent: The set of parameters used to fit a given set of low-energy observables is not
unique. Even worse, two sets of parameters, which both give the same results for the low-energy observables, do not necessarily lead to the 
same results for the chiral phase boundary and the location of the critical endpoint is not necessarily the same~\cite{Stephanov:2007fk}.

In the past few years, quite some progress has been made in connecting the low-energy regime described by quark-meson
dynamics with the dynamics at high temperatures, see e. g. Refs.~\cite{Meisinger:1995ih,Pisarski:2000eq,Fukushima:2003fw,Braun:2003ii,Megias:2004hj,Ratti:2005jh,Sasaki:2006ww,Schaefer:2007pw}. 
Such improved models for a description of the QCD dynamics at finite temperature and density are mostly based on the inclusion of a Polyakov-Loop
potential extracted from lattice QCD results. By this means, the treatment of the gauge-field dynamics has been outsourced while
the less problematic quark-meson dynamics are treated self-consistently within the framework. 
Although all of these approaches provide us with a better understanding of the thermodynamics of QCD at low and high temperatures, 
they cannot get rid of the parameter dependence of the results. In addition, the back-reaction 
of the quark-dynamics on the gauge-field dynamics in terms of the Polyakov-Loop has not yet been fully taken into account. 
For a quantitative description of the QCD phase boundary, however, not only the gauge-field dynamics need to be taken into account:
The fluctuations of the Goldstone modes and the radial mode beyond the mean-field approximation also play an important role at
the phase boundary, in particular with respect to a better description of the susceptibilities in QCD with physical pion masses.

In this paper, we discuss a functional Renormalization Group (RG) approach to the QCD phase boundary.
The chiral phase boundary of the quark-meson model (bosonized NJL model) with two degenerate quark flavors 
has been studied in the local potential approximation using a functional RG approach in Ref.~\cite{Schaefer:2004en}.
The advantage of the approach presented in this work is that it allows not only for dynamical chiral symmetry breaking 
triggered by gluodynamics but also provides access to the infrared domain of QCD dominated by pions.
Our approach is based on ground-breaking work done by H.~Gies and C.~Wetterich, see Refs.~\cite{Gies:2001nw,Gies:2002hq}.  
There it has been shown for vanishing temperature and quark chemical potential that both the regime dominated by Goldstone modes 
and the perturbative QCD regime dominated by quark-gluon dynamics can be conveniently linked without fine-tuning using the
functional RG. In Refs.~\cite{Braun:2005uj,Braun:2006jd}, the chiral 
phase boundary in the plane of temperature and number of quark flavors has been computed by studying quark-gluon 
dynamics using the functional RG. 
The strategy of the latter papers was to determine for which temperatures and number of quark flavors QCD remains in 
the chirally symmetric regime and thereby implicitly extracting the phase transition temperatures. In contrast,
this paper aims to set the stage for studies of the QCD phase boundary with two and three quark flavors
including the possibility to study the low-temperature regime and the order of the chiral phase transition.
To this end, we use the approach discussed in Refs.~\cite{Gies:2001nw,Gies:2002hq} and combine it with the 
findings of Refs.~\cite{Braun:2005uj,Braun:2006jd}: Our strategy is to follow the RG flow starting at high momentum 
scales $(p\sim M_Z)$ down to the deep infrared regime which is dominated by the 
dynamics of Goldstone modes. This allows us to get rid of the unwanted ambiguity in the parameter-space as it is present in
NJL-type models. Our results for the phase boundary will depend on only a single input parameter,
namely the value of the strong coupling $\alpha_s$ at the initial RG scale. By this means, the scale is set unambiguously in our calculations
and the values of all dimensionful quantities, such as the constituent quark mass or the chiral phase transition temperature, are eventually determined 
by the choice of the initial value of the strong coupling only.

The paper is organized as follows: In Sect.~\ref{Sec:RGflowSetup}, we give a discussion of the technical details of our functional RG 
approach for a study of the QCD phase boundary. Our results for the chiral phase boundary at small quark chemical potentials for 
QCD with one quark flavor including a comparison to lattice QCD results are then discussed in Sect.~\ref{Sec:Results}. 
We also discuss the possibility of merging our work with recent studies of the deconfinement phase-transition 
in pure Yang-Mills theory using functional RG methods~\cite{Braun:2007bx,Marhauser2008}.
Our concluding remarks, including a discussion of future extensions, are presented in Sect.~\ref{Sec:Conclusions}. 
\section{RG flow of the effective action}\label{Sec:RGflowSetup}
Throughout this paper we work in $d=4$ dimensional Euclidean space and employ 
the following ansatz for the effective action for our study of the phase diagram of 1-flavor QCD:
\be
\Gamma &=& \int d^4 x\,\left\{ Z_{\psi}\bar{\psi}\left(\I D\!\!\!\!\slash[A]\,+\I\gamma_0\mu\right)\psi + \frac{\bar{\lambda} _{\sigma}}{2}[(\bar{\psi}\psi)^2 - (\bar{\psi}\gamma _5 \psi)^2] 
+ \frac{1}{2}Z_{\phi}\left(\partial_{\mu}\Phi\right)^2 + U(\Phi^2)\right.\nn\\
&& \qquad\qquad+\left. \frac{\bar{h}}{\sqrt{2}}(\bar{\psi}(\vec{\tau}\cdot\Phi)\psi) + \frac{Z_F}{4}F_{\mu\nu}^a F_{\mu\nu}^a +\frac{1}{2\xi}({D}_{\mu}[\bar{A}] a_{\mu}^a)^2\right\}
+\Gamma_{\text{gauge}}\,,\label{Eq:CompleteAction}
\ee
where $D^{ij}_{\mu}=\partial _{\mu}\delta ^{ij} -\I \bar{g}T_a^{ij}A_{\mu}^a$, with $T_a$ being the hermitean gauge-group generators of the gauge
group in the fundamental representation. We have introduced the shorthand $(\bar{\psi}\psi)=\bar{\psi}^{i}\psi_{i}$
for the color indices. In the gauge sector we have included a background gauge fixing term, $\xi$ being the gauge-fixing parameter. 
We split up the gauge field into a background field $\bar{A}_{\mu}$ and a fluctuation field $a_{\mu}$, i. e. $A_{\mu}=\bar{A}_{\mu}+a_{\mu}$.
The term $\Gamma _{\text{gauge}}$ contains the ghost sector and possible higher-order
gluonic operators. We shall discuss this part of the truncation in more detail in Sec.~\ref{Sec:RunningGaugeCoupling}. The scalar fields
are combined in the $O(2)$ vector $\Phi^T=(\Phi_1,\Phi_2)$ and we have used $\vec{\tau}=(\gamma_5, \I\cdot \mathbf{1}_d)$ in order
to define the Yukawa interaction. The initial conditions for the various couplings in Eq.~\eqref{Eq:CompleteAction} at 
the ultraviolet (UV) scale $\Lambda$ are chosen such that the initial effective action is given by the (classical) QCD action functional:
\be
\Gamma _{k=\Lambda}=\int d^4 x \left\{\frac{1}{4}F_{\mu\nu}^{a}F_{\mu\nu}^{a} + \bar{\psi} \I\slash\!\!\!\!D \psi\right\}\,,\label{eq:UVaction}
\ee
see also the discussion in Sec.~\ref{SubSec:FixedPoint}. While the inclusion of higher gluonic operators is discussed in Sec.~\ref{Sec:RunningGaugeCoupling}, 
the quark-meson part of our truncation is built around the standard mean-field ansatz of the effective action (large $N_c$ ansatz; standard NJL model ansatz), 
i.~e. $(Z_{\phi}(k\!=\!\Lambda) = 0, \partial_t Z_{\phi}=0;Z_{\psi}(k\!=\!\Lambda)= 1,\partial _t Z_{\psi}= 0)$), which has been used extensively for studies of the 
QCD phase diagram, see e.~g. Refs.~\cite{Meyer:2001zp,Buballa:2003qv}. Such an ansatz underlies also most of the recent (P)NJL studies of hot and dense 
QCD, see e.~g. Refs.~\cite{Fukushima:2003fw,Megias:2004hj,Ratti:2005jh,Sasaki:2006ww,Schaefer:2007pw}.

In the present paper, we are aiming at a dynamical connection of the high- and low-momentum regime of QCD. Therefore we have to go beyond the zeroth-order 
ansatz (standard NJL-model ansatz) for the effective action. In the following we systematically extend this zeroth-order ansatz in two directions, namely in derivatives 
and $n$-point functions $\Gamma ^{(n)}$ where $n$ defines the number of legs. In order to study spontaneous symmetry breaking indicated by a non-trivial minimum 
of the order-parameter potential $U(\Phi^2)$ in Eq.~\eqref{Eq:CompleteAction}, we expand the potential in powers of $\Phi^2$ resulting in RG flow equations for the 
mesonic $n$-point functions, see Sec.~\ref{Sec:EffPot} for details. The quality of such a systematic expansion of the effective potential $U(\Phi^2)$ has been studied 
quantitatively in Refs.~\cite{Tetradis:1993ts,Papp:1999he} and is well under control. On the other hand, we perform a derivative expansion which renders the $n$-point 
functions momentum-dependent. The latter is indispensable for a connection of the high- and low-momentum regime of QCD.

Next to the zeroth-order approximation one needs to include kinetic terms for the meson fields in the truncation. The minimal truncation which allows for an inclusion of 
meson loops is given by the so-called  Local Potential Approximation (LPA), i.~e.~$(Z_{\phi}(k\!=\!\Lambda) = 1, \partial_t Z_{\phi}=0;Z_{\psi}(k\!=\!\Lambda)= 1,\partial _t Z_{\psi}= 0)$.
This truncation has been used, e.~g., in Refs.~\cite{Schaefer:1999em,Braun:2003ii} for a study of the quark-meson model at finite temperature and density. It indeed turns 
out that the LPA represents already a major improvement with respect to the quality of the critical exponents; the quality of critical exponents can be considered as a measure 
of how good the dynamics at the phase transition are captured. In the present paper, we go also beyond this approximation and allow for a running of the wave-function 
renormalizations of the quark and meson fields\footnote{We would like to remark that a truncation with $(Z_{\phi}(k\!=\!\Lambda) \to 0, \partial_t Z_{\phi}\neq0;Z_{\psi}(k\!=\!\Lambda)= 1,\partial _t Z_{\psi}= 0)$ represents the lowest order in the derivative expansion which allows for a dynamical connection of the high- and low-momentum regime of QCD, see discussion 
in Sec.~\ref{sec:RebosCoup} and Refs.~\cite{Gies:2001nw,Gies:2002hq}.}, i.~e. $(Z_{\phi}(k\!=\!\Lambda) \to 0, \partial_t Z_{\phi}\neq 0;Z_{\psi}(k\!=\!\Lambda)= 1,\partial _t Z_{\psi}\neq 0)$. 
Such a truncation renders the involved vertices momentum-dependent and improves the quality of the results as it can be read off from, e.~g., the quality of the 
critical exponents\footnote{The derivative expansion can be continued systematically by, e.~g. including terms of the form $Y_k(\Phi\partial_{\mu} \Phi)^2$ in our 
truncation, see e.~g. Ref.~\cite{VonGersdorff:2000kp}.}, see~e.~g. Refs.~\cite{Tetradis:1993ts,Berges:1997eu,Berges:2000ew}. Aside from an extension
of a given truncation with higher-order operators, an error estimate for a given truncation can be obtained by a variation of the regulator. By this means
it was found in Ref.~\cite{Gies:2005as} that the present truncation~\eqref{Eq:CompleteAction} gives remarkably robust results when applied to a study of 
chiral symmetry breaking at vanishing temperature and chemical potential. Although we have not performed such a variation of the regulator in the present work, 
it is likely that the findings in Ref.~\cite{Gies:2005as} hold also in the present context of chiral symmetry breaking at finite temperature. This is due to the fact that
chiral symmetry breaking sets in on scales $T/k \lesssim 0.5$ as we shall see below.

We would like to point out that our truncation~\eqref{Eq:CompleteAction} is redundant since the four-fermion coupling $\bar{\lambda}_{\sigma}$ is related to the scalar 
potential $U(\Phi^2)$ and the Yukawa coupling $\bar{h}$ via a Hubbard-Stratonovich transformation. However, this redundancy can be completely lifted by applying 
"re-bosonization" techniques~\cite{Gies:2001nw,Gies:2002hq,Pawlowski:2005xe} which we will use here. This allows us to conveniently bridge the gap between quark 
and gluon degrees of freedom in the UV and mesonic degrees of freedom in the infrared (IR) regime. Moreover, re-bosonization techniques allow us to conveniently 
include momentum-dependent fermionic $n$-point functions (four-fermion interactions, six-fermion interactions, ...) up to arbitrary order in the RG flow\footnote{Note that maximal 
$n$ is related to the maximal order in our expansion of the order-parameter potential in $\Phi^2$ ($\Phi\sim\bar{\psi}\psi$) by means of the continuously performed Hubbard-Stratonovich 
transformations in the RG flow.}. In this paper, we work along the lines of Ref.~\cite{Gies:2002hq} and give only a brief discussion of the "re-bosonization"-procedure in 
Sec.~\ref{sec:RebosCoup}. 

The effective action~\eqref{Eq:CompleteAction} has a global $U_A(1)$ symmetry. The breaking of this global symmetry is associated with 
topologically non-trivial gauge configurations. For the moment, we do not include terms that break this global
symmetry, but we discuss the effect of such terms on our results in Sec.~\ref{Sec:Results}. In one-flavor QCD,
these gauge configurations play a very exposed role since they induce masslike fermion interactions which break the $U_A(1)$
symmetry~\cite{tHooft:1976fv,Shifman:1979uw,Shuryak:1981ff,Schafer:1996wv}. 
The impact of gauge-field configurations with non-trivial topology on the nature of the chiral phase transition of 2-flavor QCD is not
yet conclusively settled. In general, it is expected that such gauge-field configurations become less important with increasing 
number of quark flavors~\cite{Pisarski:1983ms}. In this respect, we expect that dropping $U_A(1)$ violating terms in
our ansatz for the effective action of 1-flavor QCD makes the phase structure more closely comparable to QCD with more than one quark flavor.

For our derivation of the RG flow equations of the couplings, we employ the Wetterich equation~\cite{Wetterich:1992yh}:
\be
\partial _t \Gamma_{k} [\chi] 
= \frac{1}{2} \STr\left\{\left[\Gamma _{k} ^{(1,1)}[\chi] + R_k\right]^{-1}
\!\cdot\!\left(\partial _t R_k\right)\right\}
\;\;\text{with}\;\; 
\Gamma _{k} ^{(1,1)}[\chi]=\fderL{\chi ^{T}}\Gamma _{k} [\chi]\fderR{\chi}\,,\label{eq:Flow_EffAction}
\ee
where $t=\ln k/\Lambda$ and $\Lambda$ is the UV cutoff. Here, $\chi$ represents a vector in field space and is defined as
\be
\chi ^T \equiv \chi ^T (-q):=\left(A_{\mu}^{T}(-q),\psi ^T (-q),\bar{\psi} (q),\Phi_1 (-q),\Phi_2 (-q)\right)\nn
\ee
and
\be
&&\qquad\quad\chi\equiv\chi(q):= \left(\begin{array}{c}
{ A_{\mu}(q)}\\
 {\psi (q)}\\
{\bar{\psi}^T(-q)}\\
{\Phi_1 (q)}\\
{\Phi_2 (q)}
\end{array}
\right)\,.\nn
\ee
Thus, $\Gamma _{k} ^{(1,1)}[\chi]$ is matrix-valued in field space and so is the regulator function $R_k$.
In this work, we employ a 3$d$ optimized regulator function which is technically advantageous for studies at
finite temperature~\cite{Braun:2003ii,Litim:2006ag,Blaizot:2006rj}. The quality of such a 3$d$ regulator in the 
limit of vanishing temperature and chemical potential has been estimated by computing critical exponents 
of $O(N)$ models~\cite{Litim:2001hk} and comparing them to those optained with an optimized regulator 
in 4$d$ space-time~\cite{Litim:2000ci,Litim:2001up}.
Details on the regularization can be found in App.~\ref{sec:thresholdfcts}. In the following we give the RG flow 
equations in a way which does not dependent on the details of our 3$d$ regularization. Reviews on and introductions to 
the functional RG and its application to gauge theories can be found in 
Refs.~\cite{Litim:1998nf,Berges:2000ew,Polonyi:2001se,Pawlowski:2005xe,Gies:2006wv}.

Decomposing the inverse regularized propagator on the RHS of Eq.~\eqref{eq:Flow_EffAction} into a field-independent 
and a field-dependent part,
\be
\Gamma _{k} ^{(1,1)}[\chi]+R_k ={\mathcal P}_k + {\mathcal F}_k\,,
\ee
we can expand the flow equation in powers of the fields:
\be
\label{eq:flowexp}
\partial_{t}\Gamma_{k}&=&\frac{1}{2}\STr\bigg\{\tilde{\partial}_{t}\ln(\mathcal{P}_k+\mathcal{F}_k)\bigg\}\nn\\
&=&\frac{1}{2}\STr\bigg\{\tilde{\partial}_{t}\left(\frac{1}{\mathcal{P}}_{k}\mathcal{F}_{k}\right)\bigg\}
\!-\!\frac{1}{4}\STr\bigg\{\tilde{\partial}_{t}\left(\frac{1}{\mathcal{P}}_{k}\mathcal{F}_{k}\right)^{2}\bigg\}
\!+\!\frac{1}{6}\STr\bigg\{\tilde{\partial}_{t}\left(\frac{1}{\mathcal{P}}_{k}\mathcal{F}_{k}\right)^{3}\bigg\}+\!\dots.
\ee
Here, $\tilde{\partial}_{t}$ denotes a formal derivative acting only on the $k$-dependence of the regulator 
function~$R_k$. The powers of $\frac{1}{\mathcal{P}}_{k}\mathcal{F}_{k}$ can be computed by simple matrix multiplications.
The flow equations for the various couplings can then be calculated by comparing the coefficients of the operators appearing 
on the RHS of Eq.~\eqref{eq:flowexp} with the couplings specified in our truncation. 
\subsection{RG flow of the effective potential}\label{Sec:EffPot}
In this section we discuss the effective potential. The RG flow of the effective potential receives contributions from
the scalar as well as the fermionic degrees of freedom:
\be
U(\Phi)=U_{B}(\Phi) + U_{F}(\Phi)\,.\label{eq:effpot}
\ee
It also depends implicitly and explicitly on the gauge degrees of freedom. The implicit dependence
affects the running of the scalar couplings whereas the explicit dependence would result in additional terms 
in Eq.~\eqref{eq:effpot}. Since we are not interested in thermodynamic quantities such as the pressure, but only in
the order parameter, we can neglect these explicit contributions here. The contribution of the scalar fields to the 
effective potential is given by
\be
U_{B}(\Phi)&=&
\frac{1}{2}T\sum_{n=-\infty}^{\infty}\int\frac{d^{3}p}{(2\pi)^{3}}\vec{p}^{\,2} (\partial _t r_{B,k})\Big\{
Z^{\perp}_{\sigma}(\omega_n,\{p_i\}) P_B(\bar{M}_{\sigma}(\Phi))\nn\\
&&\hspace*{7cm} +\, Z^{\perp}_{\pi}(\omega_n,\{p_i\}) P_B (\bar{M}_{\pi}(\Phi))
\Big\}\,,\label{eq:UBfloweq} 
\ee
where $T$ defines the temperature and $\omega_n=2\pi nT$ denotes the bosonic Matsubara frequencies. The functions 
$Z_{\sigma}^{\perp}$ and $Z_{\pi}^{\perp}$ are the wave-function renormalizations of the sigma field 
and pion field perpendicular to the heat bath. The definition of the momentum dependent boson propagator $P_B$ can 
be found in App.~\ref{app:propdef}. The masses $M_i$ of the scalar fields are in general momentum dependent, $\bar{M}_i=\bar{M}_i(p_0,\{p_i\})$, 
and given by the eigenvalues of the second derivative matrix of the potential,
\be
\bar{M}_{ij}(\Phi,p,q)=\fderL{\,(\delta\Phi^T _i (-p))} \int d^4 x\, U(\Phi + \delta\Phi) \fderR{\,(\delta\Phi _j (q))}\quad\text{with}\quad\delta\Phi^{T}
=(\delta\Phi_1^{T},\delta\Phi_2^{T}),
\ee
evaluated at the scalar background-field configuration $\Phi$. In the following, we approximate the 
full potential $U$ by a Taylor expansion in terms of the fields around the physical ground-state $\Phi _0$ up to quartic 
order\footnote{We neglect higher-order terms since we are not aiming at a high-accuracy determination of critical exponents,
where such higher-order terms have proven their importance, see e. g.~\cite{Tetradis:1993ts,Litim:2002cf,Bervillier:2007rc}}. 
Note that such a low-order expansion of the chiral order-parameter potential is incapable of describing a first-order phase 
transition. In particular, we are not able to detect the emergence a critical endpoint. However, our present work can be 
generalized straightforwardly along the lines of Refs.~\cite{Litim:1994jd} or~\cite{Schaefer:2004en} where first-order transitions have been 
studied within RG approaches.

In the regime with an $O(2)$ symmetric ground-state ($\Phi_0 =0$), we use
\be
U_{\text{sym}} (\Phi)=\frac{1}{2}m^2 \Phi^2 + \frac{\bar{\lambda}_{\phi}}{4} \Phi^4\,.\label{eq:UsymAnsatz}
\ee 
In this case, the masses of the scalar fields are given by
\be
\bar{M}_{\sigma}^2(\Phi)&=&2\frac{\partial U_{\text{sym}}}{\partial \Phi^2} +  4\Phi^2 \frac{\partial ^2 U_{\text{sym}}}{\partial \Phi^2\partial \Phi^2}= m^2 + 3{\bar{\lambda}}_{\phi}\Phi ^2\,,\\
\bar{M}_{\pi}^2(\Phi)&=&2\frac{\partial U_{\text{sym}}}{\partial \Phi^2} = m^2 + {\bar{\lambda}}_{\phi}\Phi ^2\,,
\ee
and the physical masses $M_i(\Phi=0)$ are degenerate. In the regime with spontaneously broken $O(2)$ symmetry of the 
ground-state ($\langle \Phi\rangle\equiv\Phi_0\neq 0$), we use the ansatz
\be
U_{\text{bro}} (\Phi)=\frac{\bar{\lambda}_{\phi}}{4} (\Phi^2 - \Phi _0 ^2)^2\,,\label{eq:UbroAnsatz}
\ee
which yields
\be
\bar{M}_{\sigma}^2(\Phi)&=& 2\frac{\partial U_{\text{bro}}}{\partial \Phi^2} +  4\Phi^2 \frac{\partial ^2 U_{\text{bro}}}{\partial \Phi^2\partial \Phi^2} = \bar{\lambda}_{\phi} (\Phi^2 - \Phi _0^2) 
+ 2\bar{\lambda}_{\phi} \Phi^2\,,\\
\bar{M}_{\pi}^2(\Phi)&=&2\frac{\partial U_{\text{bro}}}{\partial \Phi^2}=\bar{\lambda}_{\phi} (\Phi^2 - \Phi _0^2) \,.
\ee
for the masses of the scalar fields.

The fermionic contribution $U_{F}$ to the effective potential $U$ reads
\be
U_{F}(\Phi)
&=&2\,N_c\, T\sum_{n=-\infty}^{\infty}\int\frac{d^{3}p}{(2\pi)^{3}}\vec{p}^{\,2} (\partial _t r_{\psi,k})\Big\{
Z^{\perp}_{\psi}(\nu_n,\{p_i\}) {\mathcal P}^{(+)}(\bar{M}_{\psi})\nn\\
&& \hspace*{7cm} +\, Z^{\perp}_{\psi}(-\nu_n,\{-p_i\}) {\mathcal P}^{(-)}(\bar{M}_{\psi})
\Big\}\,,\label{eq:UFfloweq} 
\ee
where $\nu_n=(2n+1)\pi T$ denotes the fermionic Matubara frequencies and the fermion mass is given by
\be
\bar{M}_{\psi}^2\equiv\bar{M}_{\psi}^2 (\nu_n,\{p_i \})=\frac{1}{2}(\bar{h}(\nu_n,\{p_i \}))^2 \Phi^2\,.
\ee
The fermion propagators $ {\mathcal P}^{\pm}$ are defined in App.~\ref{app:propdef}. As we shall see in Sec.~\ref{subsec:DerivOfYukawa}, the wave-function 
renormalizations $Z_{\psi}^{\perp}$ and $Z_{\psi}^{\parallel}$ and the fermion mass have the property
\be
(Z^{\perp}_{\psi}(\nu_n,\{p_i\}))^{*} = Z^{\perp}_{\psi}(-\nu_n,\{p_i\}),&\quad& (Z^{\parallel}_{\psi}(\nu_n,\{p_i\}))^{*} = Z^{\parallel}_{\psi}(-\nu_n,\{p_i\})\nn\\
\text{and}\quad(\bar{M}_{\psi}(\nu_n,\{p_i \}))^{*} &=& \bar{M}_{\psi}(-\nu_n,\{p_i \}).
\ee
Thus the fermion propagator obeys
\be
({\mathcal P}^{(+)}(\bar{M}_{\psi}(\nu_n,\{p_i\})))^{*}={\mathcal P}^{(-)}(\bar{M}_{\psi}(-\nu_n,\{p_i\}))
\ee
and the fermionic contribution $U_{F}$ to the effective potential is {\it real-valued}, as it should be:
\be
U_{F}(\Phi)
&=&8 N_c\, T\,\mathbf{Re}\sum_{n=0}^{\infty}\int\frac{d^{3}p}{(2\pi)^{3}}\vec{p}^{\,2} (\partial _t r_{\psi,k})
Z^{\perp}_{\psi}(\nu_n,\{p_i\}) {\mathcal P}^{(+)}(\bar{M}_{\psi})\,.    
\ee
The fact that  $U_F$, and thus $U$, is real-valued is an important property of the effective potential, since it can then be expanded in powers of $\mu^2/(\pi^2 T^2)$. 
As a consequence, we can expand the phase boundary in powers of $\mu ^2/(\pi T)^2$ around $\mu=0$. 

The RG flow equations for the couplings $m^2$, $\lambda_{\phi}$ and the vacuum expectation value $\Phi_0$ can now be calculated by projecting the RHS of 
Eqs.~\eqref{eq:UBfloweq} and~\eqref{eq:UFfloweq} onto our ans\"atze~\eqref{eq:UsymAnsatz} and~\eqref{eq:UbroAnsatz} for the potential $U$. In order to study 
the RG flow of the potential, we introduce the following dimensionless renormalized quantities:
\be
\epsilon=\frac{m^2}{Z_{\phi}^{\perp}k^2}\,, \quad \lambda_{\phi}=\frac{\bar{\lambda}_{\phi}}{(Z_{\phi}^{\perp})^2}\,, \quad
\kappa=\frac{1}{2}\frac{Z_{\phi}^{\perp}\Phi^2_0}{k^2}\,,\quad h^2=\frac{\bar{h}^2}{Z_{\phi}^{\perp}(Z_{\psi}^{\perp})^2}\,.
\ee
For the symmetric regime, we then find (with $v_3=\frac{1}{8\pi^2}$)
\be
\partial _t \epsilon &=& (\eta_{\phi}^{\perp} -2)\epsilon - 8 v_{3} \lambda_{\phi}\, l_1 ^{(B),(4)}(\tilde{t},\epsilon;\eta_\phi ^{\perp})
+ 8 N_c v_{3} h^2 \, l_1 ^{(F),(4)}(\tilde{t},0,\tilde{\mu};\eta_\psi^{\perp})\,,\\
\partial _t \lambda_{\phi} &=& 2\eta_{\phi}^{\perp}\lambda_{\phi} +  20 v_{3} \lambda_{\phi}^2 \, l_2 ^{(B),(4)}(\tilde{t},\epsilon;\eta_\phi^{\perp})
- 8 N_c v_{3} h^4 \, l_2 ^{(F),(4)}(\tilde{t},0,\tilde{\mu};\eta_\psi^{\perp})\,,
\ee
where $\tilde{t}=T/k$ and $\tilde{\mu}=\mu/k$ denote the dimensionless temperature and dimensionless chemical potential. The anomalous dimensions $\eta_{\phi} ^{\perp}$ 
of the scalar and $\eta_{\psi}^{\perp}$ of fermion field are given by
\be
\eta_{\phi}^{\perp}=-\partial_t \ln Z^{\perp}_{\phi}\qquad\text{and}\qquad \eta_{\psi}^{\perp}=-\partial_t \ln Z^{\perp}_{\psi}\,.
\ee
For the regime with broken O(2) symmetry in the ground-state, we find the following flow equations:
\be
\partial _t \kappa &=& -(\eta_{\phi}^{\perp} + 2)\kappa + 6 v_{3} \, l_1 ^{(B),(4)}(\tilde{t},2\kappa\lambda_{\phi};\eta_\phi^{\perp})
+ 2 v_{3} \, l_1 ^{(B),(4)}(\tilde{t},0;\eta_\phi^{\perp})\nn\\
&&\hspace*{7cm} - 8 N_c v_{3} \frac{h^2}{\lambda_{\phi}} \, l_1 ^{(F),(4)}(\tilde{t},\kappa h^2,\tilde{\mu};\eta_\psi^{\perp})\,,\\
\partial _t \lambda_{\phi} &=& 2\eta_{\phi}^{\perp}\lambda_{\phi} +  18 v_{3} \lambda_{\phi}^2 \, l_2 ^{(B),(4)}(\tilde{t},2\kappa\lambda_{\phi};\eta_\phi^{\perp})
+  2 v_{3} \lambda_{\phi}^2 \, l_2 ^{(B),(4)}(\tilde{t},0;\eta_\phi^{\perp})\nn\\
&&\hspace*{7cm} - 8 N_c v_{3} h^4 \, l_2 ^{(F),(4)}(\tilde{t},\kappa h^2,\tilde{\mu};\eta_\psi^{\perp})\,.
\ee
The threshold functions are defined in App.~\ref{sec:thresholdfcts} and can be represented as Feynman diagrams associated with purely bosonic and fermionic loops 
involving the corresponding full propagators. The regulator dependence of the flow equations is absorbed into these functions.
\subsection{RG Flow of the Yukawa coupling}\label{subsec:DerivOfYukawa}
We now turn to the calculation of the flow equation of the Yukawa coupling. Expanding the flow equation up to second order in the fermionic fields, we find at 
$T=0$ and $\mu=0$~\cite{Jungnickel:1995fp}:
\be
&&\delta \Gamma ^{(2)}_{k,\bar{\psi}\psi}=
\frac{1}{4}\int\frac{d^4 q}{(2\pi)^4}\,\tilde{\partial}_t 
\Big\{ \left[\bar{h}(\frac{q_0+Q_0}{2},\{\frac{q_i+Q_i}{2}\})\right]^2\bar{\psi}P^{(+)}_{\psi}(\bar{M}_{\psi})P_{B,\sigma}(Q_0-q_0,\{Q_i - q_i\}) \psi\nn\\
&&\hspace*{3cm} -\; \left[\bar{h}(\frac{Q_0-q_0}{2},\{\frac{Q_i-q_i}{2}\})\right]^2\bar{\psi}P^{(-)}_{\psi}(\bar{M}_{\psi})P_{B,\sigma}(Q_0 + q_0,\{Q_i + q_i\}) \psi 
\Big\}\quad \nn\\
&&\quad\qquad\qquad - \quad (P_{B,\sigma}\;\rightarrow\;P_{B,\pi})\,,\label{Eq:flow}
\ee
where $Q_{\mu}$ denotes the four-momenta of an incoming fermion. The flow equation for the Yukawa coupling $\bar{h}(Q_0,\{Q_i\})$ is 
obtained from this expression by projecting it onto the operator $\frac{1}{\sqrt{2}}(\bar{\psi}\,\vec{\tau}\cdot\Phi\,\psi)$. 

For finite temperature $T$, the integral in $q_0$-direction becomes a sum over Matsubara frequencies. Let us now discuss this integral/sum in 
Euclidean time direction in Eq.\eqref{Eq:flow}. Since we use a 3$d$ regulator, we can study the integral/sum in Euclidean time direction 
without discussing details of the regulator function or of the integration in spatial directions.

As a first approximation~\cite{Jungnickel:1995fp}, we set $q_0 \to Q_0$ and $q_i \to Q_i$ in the argument of the fermion mass and the Yukawa coupling 
on the RHS of Eq.~\eqref{Eq:flow} and take then the limit of vanishing spatial external momenta $Q_i\to 0$. As we shall discuss further in the next subsection, 
we also neglect a possible difference between $Z_{\psi,B}^{\perp}$ and $Z_{\psi,B}^{\parallel}$, thus $Z_{\psi,B}\equiv Z_{\psi,B}^{\perp}=Z_{\psi,B}^{\parallel}$. 
We then obtain the following expression for $\delta \Gamma ^{(2)}_{k,\bar{\psi}\psi}$:
\be
&&\delta \Gamma ^{(2)}_{k,\bar{\psi}\psi}=\frac{(\bar{\psi}\vec{\tau}\cdot\Phi\psi)}{2\sqrt{2}}\int\frac{d^{3}q}{(2\pi)^{3}}\,\tilde{\partial}_t
\,\delta \tilde{\Gamma} ^{(2)}_{k,\bar{\psi}\psi}(\{q_i\},Q_0,0)\; + \; (P_{B,2}\;\rightarrow\;P_{B,1})\,,\label{Eq:deltaGamma1}
\ee
with
\be
&&Z_{\phi}^{-\frac{1}{2}}Z_{\psi}^{-1}\left(\delta \tilde{\Gamma} ^{(2)}_{k,\bar{\psi}\psi}(\{q_i\},Q_0,0)\right)\label{Eq:deltaGamma4}\\
&&=\left[h(Q_0,0)\right]^3\int \frac{dq_0}{2\pi}
\frac{1}
{\vec{q}^{\,2} (1+r_{\psi})^2 + (q_0  + \I\mu)^2 + M_{\psi} ^2 (Q_0,0)}
\frac{1}{\vec{q}^{\,2} (1+r_B) + (Q_0 - q_0) ^2 + M _{B,2}^2}\,\nn
\ee
and $M_{B}^2=\bar{M}_B^2/Z_{\phi}$ and $M_{\psi}^2=\bar{M}_{\psi}^2/Z_{\psi}^2$.

Let us first consider the case of finite chemical potential but zero temperature. In this case we observe that Eq.~\eqref{Eq:deltaGamma4} 
is a real number if and only if $Q_0$ is zero. This can be seen by writing the fermion propagator as follows:
\be
\frac{1}{\vec{q}^2 (1+r_{\psi})^2 + (q_0  + \I\mu)^2 + M_{\psi} ^2 (Q_0,0)}=
\frac{\vec{q}^{\,2} (1+r_{\psi})^2 + M_{\psi} ^2 (Q_0,0) + q_0 ^2 - \mu ^2 - 2\I q_0\mu}{|\vec{q}^{\,2} (1+r_{\psi})^2 + (q_0  + \I\mu)^2 + M_{\psi} ^2 (Q_0,0)|^2}\,.
\label{Eq:deltaGamma4_1}
\ee
Thus the imaginary part of the fermion propagator is linear in $q_0$ and vanishes by integration for $Q_0\to 0$.
This means, however, that the Yukawa coupling becomes a complex number for a finite external time-like momenta $Q_0$ since in this 
case the integrand in Eq.~\eqref{Eq:deltaGamma4} is no longer symmetric in $q_0$. 

Now we switch on temperature. Since $Q_0$ is the Euclidean time component of an incoming fermion, we have $Q_0 = (2m +1)\pi T\equiv \nu _m$. 
Taking into account that the integral in Eq.~\eqref{Eq:deltaGamma4} runs over fermionic momenta ($q_0 = (2n +1)\pi T\equiv \nu_n$), we find
\be
&&Z_{\phi}^{-\frac{1}{2}}Z_{\psi}^{-1}\left(\delta \tilde{\Gamma} ^{(2)}_{k,\bar{\psi}\psi}(\{q_i\},\nu _m,0)\right)\label{Eq:deltaGamma5}\\
&&=\left[h(\nu_m,0)\right]^3 T\!\sum _{n=-\infty} ^{\infty}\!
\frac{1}
{\vec{q}^{\,2} (1\!+\! r_{\psi})^2 \!+\! (\nu_n \!+\! \I\mu)^2 \!+\! M_{\psi} ^2 (\nu_m,0)}
\frac{1}{\vec{q}^{\,2} (1\!+ \!r_B)\! +\! (\nu_m \!-\! \nu_n) ^2\! +\! M _{B,2}^2}.\nn
\ee
Since $\nu_m - \nu_n = 2(m-n)\pi T$ is effectively a bosonic Matsubara frequency, the sum over~$n$  is not symmetric in $n$ and
we find that the RHS is in general a {\it complex} number  for any given value of $m$, and so is the Yukawa coupling and the fermion mass. 
From Eq.~\eqref{Eq:deltaGamma5}, we conclude
\be
\left(h(\nu _m,0)\right)^{*} = h(-\nu _m,0)\,,
\ee
and equivalently for the fermion mass. Thus we have found that the Yukawa coupling for a given external momenta is in general complex-valued 
when evaluated at a finite quark chemical potential. This is not an issue as long as we take the full momentum dependence
of the Yukawa coupling into account, as we have argued in Sec.~\ref{Sec:EffPot}. 
In the following we shall restrict ourselves to a momentum-independent Yukawa coupling which has been successfully employed in studies of
the quark-meson model with two quark flavors at vanishing chemical potential, see e. g. Ref.~\cite{Jungnickel:1995fp,Berges:1997eu,Berges:2000ew}.
This requires care in finding a proper approximation scheme that gives us a real-valued effective potential $U$ without computing the full momentum 
dependence of the couplings. The idea for constructing such a scheme is to expand the theory around the limit $\pi T/k \to 0$ of vanishing external momenta. 
This might appear dangerous for $k \lesssim \pi T$, but what comes to our rescue is the fact that $\pi T/k \lesssim 1/2$ above the scale $k_{\chi SB}$ at
which QCD enters the chirally broken regime. For scales $k<k_{\chi SB}$, the fermions acquire a mass due to the presence of a quark condensate and therefore 
the fermionic contributions to the flow decouple rapidly anyway. Thus, once chiral symmetry is broken, our approximation in the fermionic subsector should not influence 
our results much. For the purpose of implementing this truncation scheme, we introduce the following dimensionless quantities:
\be
x^2=\frac{\vec{q}^{\,2}}{k^2}\,,\quad\tilde{\nu}_n=\frac{\nu_n}{k}\quad\text{and}
\quad m^2_{\psi,B} = \frac{M_{\psi,B}^2}{k^2}\,.
\ee
Now we rewrite Eq.~\eqref{Eq:deltaGamma5} in terms of dimensionless propagators:
\be
&&Z_{\phi}^{-\frac{1}{2}}Z_{\psi}^{-1}\left(k\, \delta \tilde{\Gamma} ^{(2)}_{k,\bar{\psi}\psi}(\{q_i\},\tilde{\nu} _m,0)\right)\label{Eq:deltaGamma7}\\
&&\;=\left[h(\tilde{\nu}_m,0)\right]^3 \tilde{t}\sum _{n=-\infty} ^{\infty}
\frac{1}
{x^2 (1\!+\! r_{\psi})^2 \!+\! (\tilde{\nu}_n \! +\! \I\tilde{\mu})^2 \! +\! m_{\psi} ^2 (\tilde{\nu}_m,0)}
\frac{1}{x^2 (1\!+\! r_B) \!+\! (\tilde{\nu}_m\! -\! \tilde{\nu}_n) ^2\! +\! m _{B,2}^2}\,.\nn
\ee
Assuming that $\tilde{\nu} _m$ is a small parameter, we can expand the boson propagator in powers of~$\nu _m$:
\be
&&\frac{1}{x^2 (1+r_B) + (\tilde{\nu}_m - \tilde{\nu}_n) ^2 + m _{B}^2}\nn\\
&&\qquad\qquad\qquad=\frac{1}{x^2 (1+r_B) + \tilde{\nu}_n ^2 + m _{B}^2}  
\Big(1 +
\frac{2\tilde{\nu}_n \tilde{\nu}_m }{x^2 (1+r_B) + \tilde{\nu}_n ^2 + m _{B}^2}
+\mathcal{O}(\tilde{\nu}_m^2) 
\Big).
\ee
Equivalently, we expand the Yukawa coupling and the masses:
\be
h(\tilde{\nu}_m,0) &=& h + h^{(1)} \tilde{\nu}_m  + \mathcal{O}(\tilde{\nu}_m ^2)\,,\\
m_{\psi,B}(\tilde{\nu}_m,0) &=& m_{\psi,B}+  m_{\psi,B}^{(1)}\tilde{\nu}_m  + \mathcal{O}(\tilde{\nu}_m ^2)\,.
\ee
Keeping only the zeroth order in these expansions, we obtain for $\delta \tilde{\Gamma} ^{(2)}_{k,\bar{\psi}\psi}(\{q_i\},\tilde{\nu} _m,0)$:
\be
&&Z_{\phi}^{-\frac{1}{2}}Z_{\psi}^{-1}\left(k\, \delta \tilde{\Gamma} ^{(2)}_{k,\bar{\psi}\psi}(\{q_i\},\tilde{\nu} _m,0)\right)\nn\\
&&\quad= h^3 \tilde{t}\,\sum _{n=-\infty} ^{\infty}
\frac{1}
{x^2 (1+r_{\psi})^2 + (\tilde{\nu}_n + \I\tilde{\mu})^2 + m_{\psi} ^2}
\frac{1}{x^2 (1+r_B) + \tilde{\nu}_n ^2 + m _{B,2}^2}\,.
\label{Eq:deltaGamma8}
\ee
We observe that this expression is a real number for all $\tilde{\mu}$. Thus a Taylor expansion of this expression around $\tilde{\mu}=0$ 
generates only terms with even powers in $\tilde{\mu}$. 

Inserting~Eq.~\eqref{Eq:deltaGamma8} into Eq.~\eqref{Eq:deltaGamma1} and incorporating the gluons
in the same way as discussed here for the scalar fields, we obtain the final result for the flow of the Yukawa coupling:
\be
\partial _t h^2 &=&(\eta_{\phi}\! +\! 2\eta_{\psi})h^2 
\!-\! 4 v_3 h^4
\Big\{
L_{1,1} ^{(FB),(4)}(\tilde{t},\kappa h^2,\tilde{\mu},m_{\pi}^2;\eta_{\psi},\eta_{\phi})\! -\!  L_{1,1} ^{(FB),(4)}(\tilde{t},\kappa h^2,\tilde{\mu},m_{\sigma}^2;\eta_{\psi},\eta_{\phi})\Big\}  \nn\\
&&\hspace*{-0.8cm} -32 v_3 g^2 h^2 C_2(N_c)
\Big\{
L_{1,1} ^{(FB),(4)}(\tilde{t},\kappa h^2,\tilde{\mu},0;\eta_{\psi},\eta_{F}) \! -\! \frac{1\!-\!\xi}{3}
{\mathcal L}_{1,1} ^{(FB),(4)}(\tilde{t},\kappa h^2,\tilde{\mu},0;\eta_{\psi},\eta_{F}) \Big\}\,,
\ee
where $\eta_F=-\partial_t \ln Z_F$. Moreover, we have $(m_{\pi}^2=\epsilon,m_{\sigma}^2=2\kappa\lambda_{\phi})$ in broken 
regime and $m_{\pi}^2=m_{\sigma}^2=\epsilon$ in the symmetric regime. 
The threshold functions associated with triangle diagrams are defined in App.~\ref{sec:thresholdfcts}. We have checked that our 
results agree with those in Ref.~\cite{Gies:2002hq} in the limit $T\to 0$ and $\mu \to 0$ if we use a four-dimensional regulator function.
\subsection{RG Flow of the wave-function renormalizations}
The flow equations for the fermionic wave-function renormalization can be extracted from the RHS of Eq.~\eqref{Eq:flow} 
along the lines of the calculation of the Yukawa coupling. Projecting Eq.~\eqref{Eq:flow} onto $\bar{\psi}(-\gamma_i Q_i)\psi$ and then taking 
the limit $Q_0\to 0$ and $Q_i\to 0$, we obtain the flow of $Z^{\perp}_{\psi}$:
\be
\eta_{\psi}^{\perp}&=&\frac{4v_3}{3}C_2 (N_c) g^2\Bigg\{
4\, \mathcal{M}^{(FB),(4)}_{1,2} (\tilde{t},\kappa h^2,\tilde{\mu},0,0) -8(1-\xi)\tilde{\mathcal{N}}^{(FB),(4)}_{1,1,1}(\tilde{t},\kappa h^2,\tilde{\mu},0,0) \nn\\
&&\hspace*{3cm}+\frac{12}{5}(1-\xi)\left(\mathcal{N}^{(FB),(4)}_{1,2}(\tilde{t},\kappa h^2,\tilde{\mu},0,0) 
+ \tilde{\mathcal{N}}^{(FB),(4)}_{1,1,2}(\tilde{t},\kappa h^2,\tilde{\mu},0,0)  \right)\Bigg\}\nn\\
&&+ \frac{4v_3}{3}h^2 \Bigg\{
\left(\mathcal{M}^{(FB),(4)}_{1,2} (\tilde{t},\kappa h^2,\tilde{\mu},m_{\sigma}^2,0) + \mathcal{M}^{(FB),(4)}_{1,2} (\tilde{t},\kappa h^2,\tilde{\mu},m_{\pi}^2,0) \right)\Bigg\}\,,
\ee
where the corresponding threshold functions can be found in App.~\ref{sec:thresholdfcts}. Note that we do not display the dependence of the
threshold functions on the anomalous dimensions $\eta^{\perp}_{\phi}$ of the scalars, $\eta^{\perp}_{\psi}$ of the fermions and $\eta _F$
of the gluons for brevity, but we take it into account in the numerical evaluation of the flow equations. We find agreement with the equation for $\eta_{\psi}$ 
in Ref.~\cite{Gies:2002hq} in the limit $T\to 0$ and $\mu \to 0$ if we use a four-dimensional regulator function.
A flow equation for $Z^{\parallel}_{\psi}(Q_0,\{Q_i\})$ could in principle be obtained in the same manner by projecting
Eq.~\eqref{Eq:flow} onto~$\bar{\psi}(-\gamma_0 Q_0)\psi$.

The derivation of the scalar wave-function renormalization can be performed along the lines of Refs.~\cite{Tetradis:1993ts,Jungnickel:1995fp,Gies:2002hq}.
Projecting the flow equation onto $\phi \vec{p}^{\,2}\phi$ and taking the limit of $Q_0=0$ and $Q_i=0$ for the external momenta, we find the 
wave-function renormalization~$Z_{\phi}^{\perp}$:
\be
\eta_{\phi}^{\perp}&=&\frac{16 v_3}{3}\kappa \lambda_{\phi} ^2\, {\mathcal M}_{2,2}^{(B),(4)}(\tilde{t},m_{\sigma}^2,m_{\pi}^2;\eta_{\phi}^{\perp})
+\, \frac{40 v_3}{9}N_c h^2\, {\mathcal M}_{4}^{(F),(4)}(\tilde{t},\kappa h^2,\tilde{\mu};\eta_{\psi}^{\perp}) \nn\\
&&\qquad\qquad +\, \frac{16 v_3}{3}N_c  h^4  \kappa\,{\mathcal M}_{2}^{(F),(4)}(\tilde{t},\kappa h^2,\tilde{\mu};\eta_{\psi}^{\perp}) \,.
\ee
The threshold functions are defined in App.~\ref{sec:thresholdfcts}. For vanishing temperature and quark chemical potential, we 
find that the equation for $\eta_{\phi}^{\perp}$ agrees with the equation for $\eta_{\phi}$ provided we employ a four-dimensional
regulator function.

Here and in the following we neglect that $Z_{\phi}^{\parallel}\neq Z_{\phi}^{\perp}$ and $Z_{\psi}^{\parallel}\neq Z_{\psi}^{\perp}$
and work in the approximation $Z_{\phi}^{\parallel}= Z_{\phi}^{\perp}$ and $Z_{\psi}^{\parallel}= Z_{\psi}^{\perp}$. For our purposes,
this is justified since we are only interested in chiral symmetry breaking in QCD but not in a calculation of thermodynamical
quantities above $T_c$. For scales $k>k_{\chi SB}$, we have $Z_{i}^{\parallel}\approx Z_{i}^{\perp}$ since $T/k < 1$. 
For scales $k\ll T$, we approach the  three-dimensional limit and the RG flow is driven only by the lowest Matsubara modes. Since the
lowest Matsubara frequency for the bosons is zero, the dependence of the propagators on $Z_{\phi}^{\parallel}$ drops out, see App.~\ref{app:propdef}
for our definition of the propagators. In case of the fermions the lowest Matsubara frequency is proportional to the temperature $T$. Therefore
the fermions effectively decouple from the RG flow for $k\ll T$ and our approximation $Z_{\psi}^{\parallel}\neq Z_{\psi}^{\perp}$ 
hardly affects the RG flow. Overall, the distinction of $Z_{i}^{\parallel}$ and $Z_{i}^{\perp}$ plays only a quantitatively important role for temperatures
$T>T_c$: There the mid-momentum regime is not protected by a dynamically generated mass gap from chiral symmetry 
breaking and modes with $k\sim T$ can actually probe the difference between $Z_{i}^{\parallel}$ and $Z_{i}^{\perp}$.
\subsection{RG Flow of the four-fermion interaction}
\label{Subsec:RGFlowFourFermi}
The flow equation for the four-fermion interaction $\bar{\lambda}_{\sigma}$ can be obtained as well by projecting the expansion~\eqref{eq:flowexp} 
of the RG flow equation onto our ansatz~\eqref{Eq:CompleteAction} for the effective action. In anticipation of what follows, we note that we only need 
to take contributions arising from the fourth order term in the expansion~\eqref{eq:flowexp} into account. These are contributions 
from so-called one-particle irreducible (1PI) "box"-diagrams. As we shall see in Sec.~\ref{sec:RebosCoup},  we do not need to compute 1PI four-fermion 
self-interaction diagrams ($\sim \bar{\lambda}_{\sigma}^2$) and so-called 1 PI "triangle"-diagrams ($\sim \bar{\lambda}_{\sigma}\bar{g}^2$ 
and $\sim \bar{\lambda}_{\sigma}\bar{h}^2$), even though these diagrams would contribute to the RG flows of the four-fermion couplings in
a non-rebosonized study~\cite{Jaeckel:2003uz,Gies:2005as,Braun:2005uj,Braun:2006jd}. As a consequence, it is sufficient to consider 
the limit $\bar{\lambda}_{\sigma}\to 0$ on the RHS of the flow equation of $\bar{\lambda}_{\sigma}$. We find:
\be
\partial _t \bar{\lambda}_{\sigma}\Big|_{\bar{\lambda}_{\sigma}\to 0}
=\frac{Z_{\psi}^2}{k^2}\left(\beta _{\bar{\lambda} _{\sigma}}^{g^4}g^4 +\beta _{\bar{\lambda} _{\sigma}}^{h^4}h^4\right)\label{eq:fourfermionflow}
\ee
with
\be
\beta _{\bar{\lambda} _{\sigma}}^{h^4}&=&\frac{1}{2N_c}\frac{4v_{3}}{3}\Bigg(  
L^{(FB),(4)}_{1,1,1,1} (\tilde{t},\kappa h^2,\tilde{\mu},\tilde{\mu},m_{\sigma}^2,m_{\pi}^2) 
\!+\! L^{(FB),(4)}_{1,1,1,1} (\tilde{t},\kappa h^2,\tilde{\mu},-\tilde{\mu},m_{\sigma}^2,m_{\pi}^2)
\Bigg),\\
\beta _{\bar{\lambda} _{\sigma}}^{g^4}&=&-\frac{21 \left(C_2(N_c)\right)^2}{2 N_c}\frac{4v_{3}}{3}\Bigg(  
L^{(FB),(4)}_{1,1,1,1} (\tilde{t},\kappa h^2,\tilde{\mu},\tilde{\mu},0,0)
\!+\! L^{(FB),(4)}_{1,1,1,1} (\tilde{t},\kappa h^2,\tilde{\mu},-\tilde{\mu},0,0)
\Bigg).
\ee
The threshold functions can be found in App.~\ref{sec:thresholdfcts}. We do not display the dependence of the threshold functions on the anomalous dimensions of the 
corresponding fields for brevity but we take it into account in the numerical evaluation of the flow equations. 

We have chosen the same Fierz transformations in the Dirac algebra as in Refs.~\cite{Gies:2001nw,Gies:2002hq}. In the present study we discard additional four-fermion interactions 
of the type $(\bar{\psi}\gamma_0\psi)^2$ which are generated in the finite-temperature RG flows. However, such interactions are suppressed for scales $k > T$ compared to the included 
four-fermion interaction anyway. We have also neglected four-fermion interactions, such as a vector-channel, in our truncation of the effective action. Therefore
our results for the phase boundary will depend slightly on our choice of Fierz-transformation with respect to Dirac and color indices. However, it has been checked in 
Ref.~\cite{Gies:2002hq} that results for low-energy observables at zero temperature obtained in different Fierz decompositions involving a color-singlet scalar-pseudoscalar 
channel agree on the $1\%$ percent level. We would like to stress that it is possible to fully resolve such a Fierz ambiguity in larger truncations within the functional RG 
approch~\cite{Gies:2005as,Braun:2005uj,Braun:2006jd} even when "re-bosonization" techniques are applied~\cite{Gies:2001nw,Jaeckel:2002rm,Jaeckel:2003uz}.
\subsection{RG Flow of the rebosonized couplings}\label{sec:RebosCoup}
Let us now briefly discuss the so-called "re-bosonization" procedure which we apply in order to resolve
the redundancy in our ansatz for the effective action~\eqref{Eq:CompleteAction}. The redundancy originates from the
fact that a Yukawa coupling together with a bosonic potential can be transformed into a four-fermion interaction
and vice versa. In order to lift this redundancy we work along the lines of Ref.~\cite{Gies:2002hq} and allow for 
$k$-dependent scalar fields $\Phi_{1,k}$ and $\Phi_{2,k}$.  The flow equation~\eqref{eq:Flow_EffAction} changes then as 
follows~\cite{Gies:2001nw,Gies:2002hq}:
\be
\partial _t \Gamma_k=\partial _t \Gamma_k \Big|_{\Phi_k}+ \int \frac{d^4 q}{(2\pi)^4}\left( \frac{\delta\Gamma_k}{\delta\Phi_{1,k}}\partial_t \Phi_{1,k}
 + \frac{\delta\Gamma_k}{\delta\Phi_{2,k}}\partial_t \Phi_{2,k} \right)\,,
\ee
where the first term on the RHS is simply the flow equation~\eqref{eq:Flow_EffAction} for fixed fields $\Phi_{1,k}$ and $\Phi_{2,k}$ and the
second term takes care of the fact that the scalar fields change under a variation of the RG scale $k$. In the following,
we span the  RG flow of the scalar fields $\Phi_{1,k}$ and $\Phi_{2,k}$ by the corresponding field itself and a fermionic composite
operator with the same quantum numbers: 
\be
\partial _t \Phi_{1,k} & =& \frac{1}{\sqrt{2}}\left(\bar{\psi}\gamma_5 {\psi}\right)\partial_t \alpha_k + \Phi_{1,k} \partial _t \beta_k\,, \label{eq:fieldtrafo1}\\
\partial _t \Phi_{2,k} &=& \frac{\I}{\sqrt{2}}\left(\bar{\psi}\psi\right) \partial_t \alpha_k + \Phi_{2,k} \partial _t \beta_k\,.\label{eq:fieldtrafo2}
\ee
The functions $\alpha_k$ and $\beta_k$ determine the transformation of the scalar fields under the RG flow and can
be derived unambiguously from enforcing several conditions: The flow of $\bar{\lambda}_{\sigma}(q^2)$ must vanish on all scales $k$ 
and for all $q^2$, the Yukawa coupling must be momentum independent and the flow of the wave-function renormalization
of the scalar fields must obey $\partial _t Z_{\phi}(q^2=k^2)=-\eta_{\phi}Z_{\phi}$, see Refs.~\cite{Gies:2001nw,Gies:2002hq} for details.
Using the initial condition $\bar{\lambda}_{\sigma}|_{k\to\Lambda} =0$ for the four-fermion coupling at the UV cutoff scale $\Lambda$, the first condition 
ensures that no coupling $\bar{\lambda}_{\sigma}$ is generated in the RG flow. We stress that it is this "re-bosonization" procedure which allows 
us to bridge the gap between the perturbative quark-gluon regime in the UV and the regime dominated by massless Goldstone modes in 
the IR without performing any additional fine tuning. From a technical point of view, this technique allows us to include (momentum-dependent)
four-fermion interactions up to arbitrary order, provided we do not truncate our ansatz for the scalar potential $U(\Phi^2)$.

By applying the field transformations~\eqref{eq:fieldtrafo1} and~\eqref{eq:fieldtrafo2}, we modify the flow equations for scalar couplings. 
We find for the flow equations in the symmetric regime:
\be
\partial_t \epsilon &=&\partial _t \epsilon \Big|_{\Phi _k} + 2\frac{\epsilon (1+\epsilon)}{h^2}\left((1+\epsilon)Q_{\sigma} + 1\right)\left(\beta _{\bar{\lambda} _{\sigma}}^{g^4}g^4 +
\beta _{\bar{\lambda} _{\sigma}}^{h^4}h^4\right)\,,\\
\partial_t h^2&=&\partial _t h^2  \Big|_{\Phi _k} + 2\left((1+\epsilon)^2 Q_{\sigma} + 1+ 2\epsilon\right)\left(\beta _{\bar{\lambda} _{\sigma}}^{g^4}g^4 +
\beta _{\bar{\lambda} _{\sigma}}^{h^4}h^4\right)\,,\\
\partial _t \lambda_{\phi}&=&\partial _t \lambda_{\phi} \Big|_{\Phi _k} + 4\frac{\lambda_\phi}{h^2}\left(1+\epsilon\right)\left(1+(1+\epsilon)Q_{\sigma}\right)\left(\beta _{\bar{\lambda} _{\sigma}}^{g^4}g^4 +\beta _{\bar{\lambda} _{\sigma}}^{h^4}h^4\right)\,.
\ee
Similarily we obtain the following set of flow equations in the regime with broken $O(2)$ symmetry:
\be
\partial_t \kappa &=&\partial _t \kappa \Big|_{\Phi _k} + 2\frac{\kappa (1-\kappa\lambda_{\phi})}{h^2}\left((1+\kappa\lambda_{\phi})Q_{\sigma} 
+ 1\right)\left(\beta _{\bar{\lambda} _{\sigma}}^{g^4}g^4 +\beta _{\bar{\lambda} _{\sigma}}^{h^4}h^4\right)\,,\\
\partial_t h^2&=&\partial _t h^2  \Big|_{\Phi _k} + 2\left((1-\kappa\lambda_{\phi})^2 Q_{\sigma} + 1- 2\kappa\lambda_{\phi}\right)\left(\beta _{\bar{\lambda} _{\sigma}}^{g^4}g^4 +
\beta _{\bar{\lambda} _{\sigma}}^{h^4}h^4\right)\,,\\
\partial _t \lambda_{\phi}&=&\partial _t \lambda_{\phi} \Big|_{\Phi _k} + 4\frac{\lambda_\phi}{h^2}\left(1-\kappa\lambda_{\phi}\right)\left(1+(1-\kappa\lambda_{\phi})Q_{\sigma}\right)
\left(\beta _{\bar{\lambda} _{\sigma}}^{g^4}g^4 + \beta _{\bar{\lambda} _{\sigma}}^{h^4}h^4\right)\,.
\ee
The function $Q_{\sigma}$ occuring in the equations for the symmetric and the broken regime measures the suppression of the four-fermion interaction 
for large momenta, which we treat in an $s$-channel approximation. It is defined as~\cite{Gies:2001nw,Gies:2002hq}:
\be
Q_{\sigma}(T,\mu,\bar{M}_{\psi})
:=\frac{\partial _t ( \bar{\lambda}_{\sigma}(k^2,T,\mu,\bar{M}_{\psi}) - \bar{\lambda}_{\sigma}(0,T,\mu,\bar{M}_{\psi}))}{\partial _t \bar{\lambda}_{\sigma}(0,T,\mu,\bar{M}_{\psi})}\,.
\ee
Note that $Q_{\sigma}$ depends on the temperature $T$, the quark chemical potential $\mu$ and the quark mass $\bar{M}_{\psi}$.
In order to compute $Q_{\sigma}$, we would in principle need to compute the full momentum dependence of the four-fermion interaction. For simplicity, we 
do not perform an explicit computation of the momentum dependence but model it with the aid of theoretical constraints. First, we assume that $Q_{\sigma}<0$ in 
order to be consistent with unitarity at $T=0$. Second, the four-fermion interaction in the $s$ channel can be considered to be roughly point-like 
once the quarks acquire a mass. At finite temperature, the quarks acquire an additional thermal mass 
which further suppresses the momentum dependence. According to Ref.~\cite{Gies:2002hq}, we therefore model $Q_{\sigma}$ by employing a threshold function 
that captures these constraints:
\be
Q_{\sigma}(\tilde{t},\tilde{\mu},m_{\psi})=Q_{\sigma}^{0}\,\mathcal{M}^{(4),(FB)}_{1,2} (\tilde{t},m_{\psi}^2,\tilde{\mu},0,0;\eta_{\psi},\eta_{F})\,.
\ee
Here, $Q_{\sigma}^{0}$ is a negative constant at our disposal. We choose $Q_{\sigma}^{0}=-1$, but we have checked that our results for the phase 
boundary and in particular for the curvature at small chemical potential change only on the $1\%$ level when we vary $Q_{\sigma}^{0}$ from 
$Q_{\sigma}^{0}=-1$ up to $Q_{\sigma}^{0}=-0.01$.
\subsection{Running of the gauge coupling and gluonic anomalous dimension}
\label{Sec:RunningGaugeCoupling}
Finally we need to discuss the running of the gauge coupling which is one of the key ingredients of our study of the QCD phase boundary. 
From now on we restrict our discussion to Landau-gauge~QCD.

In this work, we use two ans\"atze for the running of the coupling in Landau-gauge QCD. This allows us to give a theoretical error estimate for our results. 
The first ansatz has been extensively discussed at both zero and finite temperature in Refs.~\cite{Braun:2005uj,Braun:2006jd,Gies:2002af}. 
It is based on the following truncation in the pure gluonic part of the effective action:
\be
\Gamma _k ^{\text{FE}}= \int _x \left\{ Z_k ^{(1)} F_{\mu \nu}^a
F_{\mu \nu}^a + Z_k ^{(2)} \left(F_{\mu \nu}^a
F_{\mu \nu}^a\right)^2 + \dots \right\}.\label{eq:gluon_trunc}
\ee
Such a calculation of the coupling employs the background-field formalism \cite{Abbott:1980hw} within the RG 
framework~\cite{Reuter:1993kw,Reuter:1997gx,Freire:2000bq,Freire:2000mn,Pawlowski:2001df,Litim:2002ce,Litim:2002hj,Gies:2002af,Pawlowski:2005xe}.

The ansatz~\eqref{eq:gluon_trunc} for the pure gluonic part of the 
effective action used for the determination of the running coupling includes an infinite power series of the gauge-invariant operator 
$F^a_{\mu\nu}F^a_{\mu\nu}$. The truncation includes arbitrarily high gluonic correlators projected onto their small-momentum limit and onto the 
particular color and Lorentz structure arising from powers of $F^a_{\mu\nu}F^a_{\mu\nu}$. It represents a gradient expansion in the field strength 
to arbitrary order but neglects higher-derivative terms and more complicated color and Lorentz structures. Using the background-field method, 
the $\beta$-function of the running coupling $g$ is related to the wave-function renormalization of the background field \cite{Abbott:1980hw} via
\begin{eqnarray}
 \partial_t \alpha _{\text{FE}}= \eta _{\text{FE}}\,\alpha_{\text{FE}}\qquad\text{with}\qquad
 \eta_{\text{FE}}=-\partial_t \ln Z_k ^{(1)}\equiv -\partial_t \ln Z_F^{\text{FE}}\,,
\label{eq:betadef}
\end{eqnarray}
which is a consequence of the non-renormalization of the product of the background field and the bare coupling. The coefficient of the 
first term  $Z_k ^{(1)}\equiv Z_F ^{\text{FE}}/4$ in the effective action \eqref{eq:gluon_trunc} evolves with the renormalization scale $k$ and is 
successively driven by all other operators in the action. In Refs.~\cite{Gies:2002af,Braun:2005uj}, the authors keep track of all contributions 
from the flows of the $Z_k ^{(i)}$ to the flow of the running coupling. An infrared fixed point for the running coupling
at zero temperature has been found  in Ref.~\cite{Gies:2002af} with $\alpha _{\text{FE}} ^{*}(T=0) \in [5.7,9.7]$. The uncertainty arises
from an unresolved color structure in the calculation. In the following we do not compute the running of the coupling
from the truncation~\eqref{eq:gluon_trunc} explicitly but use the results from Refs.~\cite{Braun:2005uj,Braun:2006jd} with $\alpha _{\text{FE}} ^{*}(T=0)=5.7$.

One of the main findings in Refs.~\cite{Braun:2005uj,Braun:2006jd} is that the coupling exhibits a non-trivial infrared fixed point at finite temperature which has 
been recently confirmed by Lattice QCD simulations~\cite{Cucchieri:2007ta}. In the low momentum regime, the solution of the RG equations exhibits a linear 
behavior with a slope determined by the infrared fixed point $\alpha_{3d}^{*}$ of the spatial 3d Yang-Mills theory \cite{Braun:2005uj,Braun:2006jd}: 
\be
\alpha _{\text{FE}}(k\ll T)\approx \alpha_{3d}^{*}\,\frac{k}{T} + {\mathcal O}\left(\left(\frac{k}{T}\right)^2\right)\,.
\ee
The value of the infrared fixed point is given by $\alpha_{3d}^{*}\approx 2.7$.
The actual presence  of this finite infrared fixed point is important for temperatures around the chiral phase transition while
the actual value of $\alpha_{3d}^{*}$ is of less importance for a study of the chiral phase transition temperature\footnote{The actual value
of $\alpha_{3d}^{*}$ may play an important role for the study of bulk thermodynamic quantities such as the pressure at high
temperatures.}. Indeed, the running coupling obtained from the truncation~\eqref{eq:gluon_trunc} has been successfully used to determine the
chiral phase boundary of QCD in the plane of temperature and number of massless quark flavors~\cite{Braun:2005uj,Braun:2006jd}.

Since our work relies partly on the background-field method, we would like to discuss briefly its advantages and disadvantages.
The application of the background-field method to functional RG flow equations has been proposed in \cite{Reuter:1993kw} and
further developed in~Refs.~\cite{Reuter:1993kw,Reuter:1997gx,Freire:2000bq,Freire:2000mn,Pawlowski:2001df,Litim:2002ce,Litim:2002hj,Gies:2002af,Pawlowski:2005xe}.
The background-field method provides a convenient framework for a study of gauge theories since it allows us in principle 
to construct a gauge-invariant effective action in a comparatively simple manner.
The background-gauge fixing procedure~\cite{Abbott:1980hw} together with the regularization lead to regulator-modified 
Ward-Takahashi identities (mWTI)~\cite{Reuter:1997gx,Freire:2000bq,Freire:2000mn}. Here, we only employ
an approximate solution to the flow in the gauge sector as obtained in Refs.~\cite{Braun:2005uj,Braun:2006jd}.
To be more specific, we identify the RG flows of the background field with those of the fluctuation field;
for a treatment of the difference of both we refer to Ref.~\cite{Pawlowski:2001df}.
The identification of the background and fluctuation field results in a flow which is no longer closed and which does not satisfy
all constraints from the mWTI~\cite{Reuter:1993kw,Reuter:1997gx,Freire:2000bq,Freire:2000mn,Pawlowski:2001df,Litim:2002ce,Litim:2002hj,Pawlowski:2005xe}. 
In this work, we assume that the loss of information due to the identification
of the background and the fluctuation field as well as corrections due to the mWTI are quantitatively small in the region of physical 
interest and do not severely affect our results, see Ref.~\cite{Litim:2002ce}. The advantage of our approximations is that 
we obtain a gauge-invariant approximate solution of the theory. In the following we work in Landau-deWitt gauge 
whenever the background-field method is involved\footnote{Note that Landau gauge is known to be a fixed point of the 
RG flow~\cite{Ellwanger:1995qf,Litim:1998qi}}. Since we shall also employ the coupling from lattice QCD in Landau gauge, our 
analysis of the phase boundary provides some quantitative insight into the quality of our approximations involved in the gauge-sector 
when treated within the background-field formalism.

In order to "measure" the impact of the gauge field dynamics on the QCD phase boundary and to estimate the theoretical error of our results arising from 
the truncation in the gauge sector, we also employ the running of the gauge coupling in Landau-gauge as measured on the 
Lattice in Ref.~\cite{Sternbeck:2006cg}. In Landau-gauge QCD, the running of the gauge coupling at vanishing temperature has been computed using lattice 
simulations~\cite{Bonnet:2000kw,Gattnar:2004bf,Dudal:2005na,Sternbeck:2006cg,Cucchieri:2007zm}, 
Dyson-Schwinger equations~\cite{vonSmekal:1997is,vonSmekal:1997vx,Fischer:2002eq,Fischer:2002hna}
and functional RG methods~\cite{Pawlowski:2003hq,Fischer:2004uk}. It can be defined by means of the ghost and the gluon propagator which is a consequence of the 
non-renormalization property of the ghost-gluon vertex~\cite{Mandelstam:1979xd,vonSmekal:1997is,vonSmekal:1997vx}:
\be
\alpha_{\text{Ref.}}(T=0,p^2) = \frac{\bar{g}^2}{4 \pi Z_A(T=0,p^2) Z_C^2(T=0,p^2)}\,,
\label{eq:LatticeCoup}
\ee
where $Z_{A,C}$ denotes the dressing functions of the gluon and the ghost, respectively. The momentum dependence of the dressing functions are 
characterized by a power-law behavior in the deep IR \cite{Zwanziger:2001kw}:
\be
Z_A^{\text{IR}}(T=0,p^2)= (p^2)^{-2\kappa _C}\,,\qquad Z_C^{\text{IR}}(T=0,p^2)=(p^2)^{\kappa_C}\,.
\ee
The exponents are related by the Landau-gauge sum rule in $d=4$ dimensions~\cite{Zwanziger:2001kw,Lerche:2002ep,Fischer:2006vf}. 
In this work, we have suitably amended the lattice propagators in Ref.~\cite{Sternbeck:2006cg} by their perturbative behavior in the ultraviolet regime 
and the corresponding power laws in the IR. This yields an infrared fixed point $\alpha_s(T=0)\approx 2.3$. 

For our finite temperature studies, we have 
adapted the running of the gauge coupling \eqref{eq:LatticeCoup} such that it is governed by an infrared fixed point for momenta $p\lesssim 2\pi T$ 
according to the results in Refs.~\cite{Braun:2005uj,Braun:2006jd}:
\be
\alpha _{\text{Ref.}}(p\ll T)\approx \alpha_{3d}^{*}\,\frac{p}{T} +  a_1\,\left(\frac{p}{T}\right)^2  +  a_2\,\left(\frac{p}{T}\right)^3 +\dots\,.\label{eq:LatticeCoupFiniteT}
\ee
Here we drop all higher terms and choose $\alpha_{3d}^{*}=1$ and determine $a_1$ and $a_2$ such that the coupling~\eqref{eq:LatticeCoup} 
and its derivative with respect to $p$ are connected continuously with the ansatz~\eqref{eq:LatticeCoupFiniteT} at the scale set by the
lowest non-vanishing bosonic Matsubara-mode $\omega_T = 2\pi T$. Although the actual values for $a_1$, $a_2$ and $\alpha_{3d}^{*}$ may differ 
from the values chosen here, the arising uncertainties for the QCD phase boundary can be estimated by a comparison with the results obtained 
from the coupling defined in Eq.~\eqref{eq:betadef}. In any case, the question whether the ground-state of QCD is governed by chiral symmetry
breaking or not is controlled by the running of the coupling in the mid-momentum regime ($0.5\,\text{GeV}\lesssim p \lesssim 1.5\,\text{GeV}$)
as we shall see below. In this momentum regime, we have $\alpha_{\text{Ref.}} > \alpha_{\text{FE}}$.

From now on we identify the scale $k^2$ set by the cutoff function with the momentum scale~$p^2$. This nontrivial assumption is justiÞed, because 
the regulator function which enters in the calculation of the running coupling specifies the Wilsonian momentum-shell integration 
in such a way that the RG flow of the coupling is dominated by fluctuations with momenta $p^2 \approx  k^2$. For our calculation of the phase boundary, 
we need not only the strong coupling, but also the anomalous dimension $\eta_F$ which we estimate from
\be
 \partial_t \alpha_{\text{Ref.}}(p^2\!=\! k^2) = \eta_{\text{Ref.}}\,\alpha_{\text{Ref.}}(p^2\!=\!k^2)\,.
\ee
By estimating the gauge-field contributions to $\eta_F=-\partial_t Z_F$ from $\eta_{\text{Ref.}}$ and by using $\alpha_{\text{Ref.}}$ in our calculations, 
we assume that the running of the coupling as found in Landau-gauge QCD can be identified with the running of the coupling found in Landau-DeWitt 
gauge within the background-field formalism. This means that we neglect possible differences between the RG flows of the fluctuation and 
the background field~\cite{Pawlowski:2001df,Litim:2002ce}. The two-loop running of the coupling is indeed identical for both gauges despite the 
fact that the couplings are defined differently. In the deep IR, there is qualitative agreement between the Landau gauge and the Landau-DeWitt gauge 
indicated by the presence of a non-trivial attractive IR fixed point. This suggests a deeper connection between both gauges~\cite{Braun:2007bx} and
justifies the use of both results for the coupling for our computation of the QCD phase boundary. The differences in the results can then be considered as a 
measure of the influence of the gauge-field dynamics on the phase boundary.

Finally, we need to discuss how the quarks affect the running of the strong coupling. In this work, we only use a one-loop
RG improved quark contribution to the gluonic anomalous dimension and therewith to the running of the strong coupling. This contribution 
can be straightforwardly computed from the effective action~\eqref{Eq:CompleteAction}. Using the regulator shape 
functions~\eqref{eq:optshapefct}, we obtain\footnote{The quark contribution $\eta_{\text{q}}$ to the gluonic anomalous
dimension should not be confused with the anomalous dimension associated with the quark propagator.}
\be
\eta _{\text{q}}(\tilde{t},\tilde{\mu},\kappa,h)=\frac{N_f}{\sqrt{1+\kappa h^2}}\left( 1 - \frac{1}{1 + \text{e}^{\frac{\sqrt{1+\kappa h^2}-\tilde{\mu}}{\tilde{t}}} } - 
 \frac{1}{1 + \text{e}^{\frac{\sqrt{1+\kappa h^2}+\tilde{\mu}}{\tilde{t}}} } \right)\frac{4}{3} \frac{g^2}{(4\pi) ^2}\,,
\ee
where $m_{\psi}$ denotes the dimensionless quark mass, and $\tilde{t}$ and $\tilde{\mu}$ denote the dimensionless temperature
and quark chemical potential, respectively. We recover the standard one-loop result in the limit of vanishing
quark mass, chemical potential and temperature, which is as it should be. The gluonic anomalous dimension $\eta _A$ is given by simply
adding the gluonic contribution and the quark contribution:
\be
\eta _{F}^{\text{Ref./FE}}=\eta_{\text{Ref./FE}}+\eta_{\text{q}}\,.
\ee
The running coupling of the strong coupling $g$ including the back-reactions of the quarks is then obtained by solving the differential
equation
\be
\partial _t \alpha _{\text{Ref./FE}} = \eta _{F}^{\text{Ref./FE}}\, \alpha _{\text{Ref./FE}}\,,
\ee
which is coupled to the the RG flow equations for the Yukawa coupling $h$ and the minimum of the scalar potential $\kappa$.
\section{The phase boundary of 1-flavor QCD at small chemical potentials}
\label{Sec:Results}
\subsection{Initial conditions and  fixed-point structure of QCD}
\label{SubSec:FixedPoint}
Let us now discuss our results for the phase boundary of QCD with one quark flavor. As initial conditions we use the running
coupling as measured at the Z-boson mass scale~$M_Z$, $\alpha_s(M_Z)\approx 0.118$~\cite{Bethke:2004uy}, and set 
the renormalized Yukawa coupling to $h(M_Z)=0.01$, the renormalized scalar mass to $m^2 (M_Z)\approx 5 M_Z ^2$ and the 
renormalized four-boson coupling to $\lambda_{\phi}=0$. With this choice, the effective action at the initial scale $M_Z$ is
effectively given by
\be
\Gamma _{k=M_Z}=\int d^4 x \left\{\frac{1}{4}F_{\mu\nu}^{a}F_{\mu\nu}^{a} + \bar{\psi} \I\slash\!\!\!\!D \psi\right\}\,.
\ee
Our choice of initial conditions is such that the results for the phase transition temperature depend only on the value of
the strong coupling at the initial RG scale. The choice of the value for strong coupling at the initial scale eventually determines 
the absolute values of observables such as the phase transition temperature or the constituent-quark mass. Let us briefly
recapitulate the results from Ref.~\cite{Gies:2002hq} and then generalize them to the case of finite-temperature QCD. 

The universal features of spontaneous chiral symmetry breaking in QCD can be nicely illustrated in terms of the fixed-point 
structure of the scalar couplings. This structure can be considered as the analogue of the fixed-point structure of four-fermion 
interactions in a purely fermionic language~\cite{Braun:2005uj,Braun:2006jd}. For this purpose, we define the coupling
\be
\tilde{\epsilon}=\frac{\epsilon}{h^2}=\frac{Z_{\psi}^2 \bar{m}^2}{k^2 \bar{h}^2}\,.
\ee
The definition of this coupling is simply motivated by the relation $\bar{\lambda}_{\sigma}\sim \bar{h}^2/\bar{m}^2$ between the 
four-fermion coupling $\bar{\lambda}_{\sigma}$ in the gauged NJL model and the scalar couplings $\bar{m}^2$ and $\bar{h}^2$ in a 
bosonized gauged NJL model.

\begin{figure}[t]
\includegraphics[scale=0.8,angle=0]{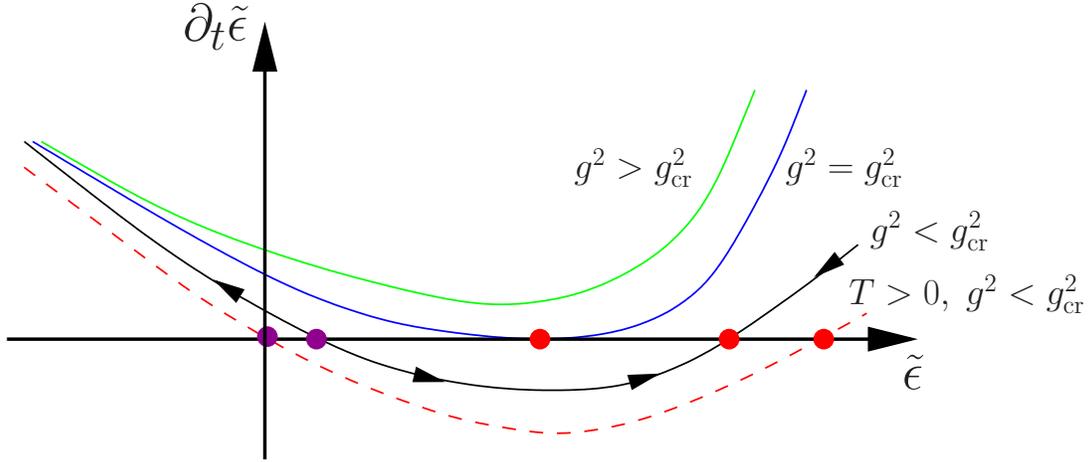}
\caption{Sketch of the $\beta$-function of the coupling $\tilde{\epsilon}$. The figure illustrates the dependence of the flow of $\tilde{\epsilon}$ 
on the gauge coupling $g$. The $\beta$-function has an UV repulsive fixed-point $\tilde{\epsilon}^{*}_1$ and IR attractive fixed-point $\tilde{\epsilon}^{*}_2$ 
if the gauge coupling $g^2$ is smaller than a critical value $g_{\text{cr}}^2$. The fixed points annihilate for $g^2=g_{\text{cr}}^2$ and no fixed points
are present for $g^2>g_{\text{cr}}^2$. The red dashed line illustrates the effect of finite temperature on the result for $g^2<g_{\text{cr}}^2$ indicated by the
black line. \label{fig:sketch}}
\end{figure}

The flow equation for the coupling $\tilde{\epsilon}$ can be straightforwardly derived from the flow of the coupling $\epsilon$
and the Yukawa coupling $h$. We find\footnote{Here we drop all arguments of the threshold functions associated with anomalous dimensions.}
\be
\beta_{\tilde{\epsilon}}\equiv\partial_t\tilde{\epsilon}&=& 8 N_c v_3 l_1^{\psi,(4)} (\tilde{t},0,\tilde{\mu}) - 8 v_3 \frac{\lambda_{\phi}}{h^2} l_1 ^{B,(4)} (\tilde{t},\epsilon)\nn\\
&& \quad -\left( 2 - 32 v_3 C_2(N_c) g^2\, \left\{ L_{1,1} ^{(FB),(4)} (\tilde{t},0,\tilde{\mu},0)-\frac{1}{3}  \mathcal{L}_{1,1} ^{(FB),(4)} (\tilde{t},0,\tilde{\mu},0)\right\}\right)\tilde{\epsilon}\nn\\
&& \quad\quad -2 \left(\beta _{\bar{\lambda} _{\sigma}}^{g^4}g^4 +\beta _{\bar{\lambda} _{\sigma}}^{h^4}h^4\right) \tilde{\epsilon}^2\,.
\label{eq:BetaEpsilonTilde}
\ee
We have neglected the anomalous dimensions for simplicity since here we are interested only in the weak coupling regime
of QCD. We find that $\beta_{\tilde{\epsilon}}$ has a UV repulsive fixed point $\tilde{\epsilon}_1^{*}$ and an IR attractive
fixed point $\tilde{\epsilon}_2^{*}$ if the gauge coupling $g$ is smaller than a critical value~$g_{\text{cr}}$, see Fig.~\ref{fig:sketch}.
For $g=g_{\text{cr}}$, the fixed points annihilate and the flow of $\tilde{\epsilon}$ is not bound by any fixed points for $g>g_{\text{cr}}$.
In the latter case, the system flows towards the regime with broken chiral symmetry characterized by a negative scalar mass 
parameter $\epsilon=m^2/k^2$. Thus, $g>g_{\text{cr}}$ represents a necessary condition for chiral symmetry breaking and the question of 
whether the QCD ground-state is chirally symmetry or not has been traced back to the strength of the gauge coupling $g$ relative to its critical value $g_{\text{cr}}$. 
We would like to mention that these considerations fully correspond to those in Refs.~\cite{Gies:2005as,Braun:2005uj,Braun:2006jd} where chiral symmetry
breaking has been studied in terms of four-fermion interactions. The critical value of $g_{\text{cr}}$ can be computed analytically 
at $T=0$ and $\mu =0$ in the limit $\epsilon \gg 1$ from Eq.~\eqref{eq:BetaEpsilonTilde}:
\be
\alpha_{\text{cr}}=\frac{g^2_{\text{cr}}}{4\pi}\approx \frac{0.27\pi}{C_2(N_c)}\qquad \Longrightarrow\qquad N_c g^2_{\text{cr}}\sim \text{const.}\text{ for }N_c\gg 1\,.
\ee
The numerical factor arises from  the evaluation of the threshold functions\footnote{The result for $\alpha_{\text{cr}}$ deviates from
the value in Ref.~\cite{Gies:2002hq} due to the choice of a 3d optimized regulator-function instead of a 4d optimized regulator-function.
The difference between both results is expected to be smaller in the case of finite temperature since the 4d and 3d optimized regulator function
coincide in the limit $T/k \to \infty$.}.  Since the influence of the Yukawa coupling and the four-boson
coupling can no longer be neglected near the chiral transition transition scale\footnote{With respect to the influence of the scalar couplings, the 
estimate for $g_{\text{cr}}$ given here does not necessarily have to agree with the value found in Refs.~\cite{Gies:2005as,Braun:2005uj,Braun:2006jd},
even if one neglects that different regulator functions have been used there. In Refs.~\cite{Gies:2005as,Braun:2005uj,Braun:2006jd}, the authors study the 
influence of gluodynamics on the quark dynamics in QCD without using rebosonization techniques. The flow of the scalar couplings is then completely 
encoded in the RG flow of the four-fermion couplings and the estimate for $g_{\text{cr}}$ within such a "pure" quark-gluon
framework should be considered to be more accurate.}, this can only serve as an estimate.

The fixed-point structure of the coupling $\tilde{\epsilon}$ allows us also to divide the initial conditions into different sets.
As in Ref.~\cite{Gies:2002hq}, the QCD starting point is given by $\epsilon_{\Lambda} =\frac{m^2}{\Lambda^2}\gg1$ since
the scalar field is then an auxiliary field at the initial scale $\Lambda=M_Z$  and its wave-function renormalization is tiny. 
Our choice for the initial conditions corresponds to a value of $\tilde{\epsilon}$ which is close to $\tilde{\epsilon}_2^{*}$. Starting the RG flow from a set of 
initial conditions obeying $\tilde{\epsilon}>\tilde{\epsilon}_2^{*}$ at the UV scale $\Lambda =M_Z$, the system flows into the fixed point $\tilde{\epsilon}_2^{*}$ 
which then controls the evolution over a wide range of scales. This explains that the IR physics at zero temperature are not sensitive 
to the choice of the initial conditions~\cite{Gies:2002hq}.

At finite temperature, the situation changes slightly. For a given value of the gauge coupling $g$, the depth of the minimum of the $\beta_{\tilde{\epsilon}}$-function
is increasing with an increase in the dimensionless temperature $T/k$ and the distance between the fixed points increases, see Fig.~\ref{fig:sketch}. 
We find
\be
\lim _{\tilde{t}\to\infty} \tilde{\epsilon}_{2} ^{*}\to\infty\,.
\ee
Thus the critical value $g_{\text{cr}}$ for the gauge coupling $g$ increases with increasing $T/k$ as well. This can be understood phenomenologically: 
The formation of a quark condensate requires stronger interactions since the quarks are thermally excited~\cite{Braun:2005uj,Braun:2006jd}.
In order to leave the result for the chiral phase boundary unaffected by the choice of the initial conditions, we have to choose the initial scale $\Lambda$
in such a way that the flow is still initially governed by the fixed-point $\tilde{\epsilon}_2^{*}$ given at $T=0$. In practice, this can be achieved by
choosing a large value for the initial scale $\Lambda$ such that $T/\Lambda \ll 0$ and $\mu/\Lambda\ll 0$ for the temperatures $T$
and quark chemical potentials $\mu$ under consideration. Our choice $\Lambda = M_Z$ translates into $T/\Lambda \sim {\mathcal O}(10^{-3})$ 
and $\mu/\Lambda \sim {\mathcal O}(10^{-3})$ for all of our numerical evaluations of the flow equations. Thus, our results for the phase boundary are 
not contaminated by the choice of the initial conditions.

Our discussion shows that the scale for all dimensionful quantities is set by the interplay between the perturbative running of the gauge coupling with its initial value,
the critical value $g_{\text{cr}}$, and the existence of the fixed point $\tilde{\epsilon}_2 ^{*}$, provided we chose QCD-like initial conditions. 
The presence of the fixed point $\tilde{\epsilon}_2 ^{*}$ ensures that the actual initial values for the scalar couplings
do not affect the absolute values for the constituent quark mass or the phase transition temperature, for example. The initial value for the gauge coupling $g$ and
its (logarithmic) running, in combination with the critical value $g_{\text{cr}}(T/k,\mu)$, then set a scale $k_{\text{cr}}(T,\mu)$. For given values of $T$ and $\mu$,
it is given by the solution of the equation
\be
g({\textstyle\frac{T}{k}},{\textstyle\frac{\mu}{k}})\stackrel{!}{=}g_{\text{cr}}({\textstyle\frac{T}{k}},{\textstyle\frac{\mu}{k}})\,.\label{eq:TcMuEstimate}
\ee
The scale $k_{\text{cr}}$ is naturally related to the scale $\Lambda_{\text{QCD}}$ at which the gauge coupling becomes large.
Thus the scale for all dimensionful quantities is set by $\Lambda _{\text{QCD}}$, independent of the initial conditions at the UV scale.
Along the lines of Ref.~\cite{Braun:2005uj,Braun:2006jd}, one can actually use Eq.~\eqref{eq:TcMuEstimate} to determine an upper bound for the chiral phase 
boundary. This is done by seeking the lowest temperature for a given $\mu$ above which Eq.~\eqref{eq:TcMuEstimate} does not have a solution anymore. 
We shall discuss this approach further in the next subsection.

Let us finally compare our approach with (P)NJL-type models. In the case of (P)NJL-type models, it is necessary to 
introduce a UV cutoff $\Lambda$ which is usually on the order of $1\,\text{GeV}$. At this UV cutoff scale the gauge degrees of freedom 
are considered to be integrated out. The cutoff can therefore be considered to have a physical meaning and is needed to define the theory. 
The strategy is then to choose the value for the four-fermion couplings (or the scalar couplings in the context of the bosonized NJL-model, i. e. 
the quark-meson model) which reproduce the values of low-energy observables, such as the pion mass, the constituent-quark mass and the pion decay 
constant. These values of the couplings are then used to compute the chiral phase boundary in the plane of temperature and quark chemical potential. 
The shortcoming of this procedure is that it does not result in unique values for the initial values of the couplings at $T=\mu=0$: 
Different sets of initial values for the couplings can reproduce the same values for the
low-energy constants under consideration but lead to different predictions for the chiral phase boundary and 
the location of the critical point\footnote{Even though the predictions from Lattice QCD simulations for the curvature of the phase boundary at small chemical 
potentials and the location of the critical endpoint do not yet agree as well~\cite{Schmidt:2006us,Stephanov:2007fk,Philipsen:2008gf}, we would like
to stress that the reasons for these differences are completely different from those encountered in the context of (P)NJL-type models.}, 
see e. g. Ref.~\cite{Stephanov:2007fk}. If we interpret (P)NJL-type models as a low-energy formulation of QCD, we can understand their parameter
ambiguity in terms of the fixed point structure of the coupling $\tilde{\epsilon}$ by considering the limit of vanishing strong coupling $g$. 
In this limit, the fixed-point value $\tilde{\epsilon}_1 ^{*}$ corresponds to the value of the inverse of the critical coupling in the NJL 
model~\cite{Gies:2001nw,Gies:2002hq}. Choosing initial conditions with $\tilde{\epsilon}_{\Lambda}< \tilde{\epsilon}_1 ^{*}$, the system is driven 
by strong four-fermion interactions and flows towards the regime with broken chiral symmetry. 
In our approach, we do not encounter such an ambiguity in the choice of the initial conditions because of the dynamically included gauge degrees of freedom
and the presence of the fixed point $\tilde{\epsilon}_2 ^{*}$. The scale for the IR physics is uniquely fixed by our choice for the initial value of
the gauge coupling $g$ and all absolute values of the observables should be interpreted in the light of this scale-fixing procedure. 

In the context of (P)NJL-type models, one might be concerned that one needs to adjust the initial values of the couplings at finite temperature. 
This concern is based on the observation that the temperature effects at the UV scale $\Lambda _{\text{NJL}}$ might be non-negligible since
$T/\Lambda_{\text{NJL}} \lesssim 0.3$ for typical UV-cutoff scales $\Lambda_{\text{NJL}}$ and temperatures used in (P)NJL-type models. 
As we have argued above, this problem is not present in our approach. 

Despite the nice features of our RG approach, the present truncation is by no means complete, in particular 
with respect to additional operators which affect the mid-momentum regime where the transition to the regime with broken chiral symmetry takes place. In this
regime, the flow is sensitive to higher-order operators owing to strong coupling. Examples we have in mind here are the inclusion of the Polyakov-Loop
in the way proposed in a RG framework for pure Yang-Mills theory in Refs.~\cite{Braun:2007bx,Marhauser2008}, or the inclusion
of operators associated with instanton effects. The latter are expected to be particularly important for 1-flavor QCD. The systematic errors
arising from neglecting such operators are discussed in the next subsection. In any case, we think that our approach is promising and 
sets the stage for future works in this direction. 
\subsection{Results for the curvature of the phase boundary of 1-flavor QCD}
\begin{figure}[t]
\includegraphics[scale=1,angle=0]{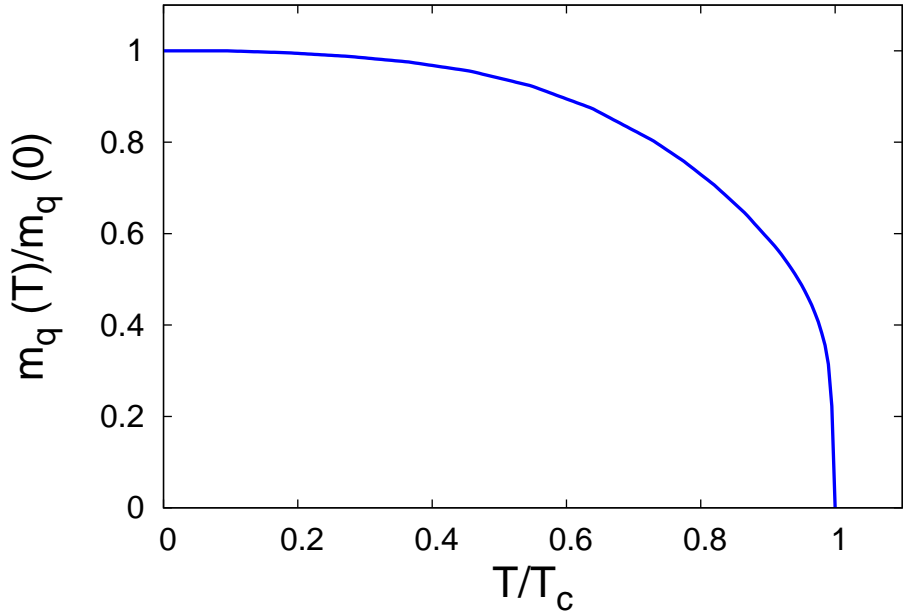}
\caption{Temperature dependence of the quark mass for vanishing chemical potential as obtained from a calculation employing 
$\alpha_{\text{Ref.}}$. The quark mass goes continuously to zero at the critical temperature indicating
a second order phase transition.\label{Fig:QuarkMassMu0}}
\end{figure}
Let us now discuss our numerical results for the (chiral) phase boundary of QCD with one quark flavor. As a first result, in Fig. ~\ref{Fig:QuarkMassMu0}
we show the temperature dependence of the quark mass for vanishing quark chemical potential. The quark mass
is related to the order parameter $\kappa$ by $m_{\psi} ^2 =\kappa h^2$ and tends to zero continuously for increasing temperature, 
which indicates a second order phase transition. This is expected since so far we have neglected $U_A(1)$ symmetry breaking terms in our study.
The results shown are obtained from the calculation employing the strong coupling $\alpha_{\text{Ref.}}$.
For the quark mass at $T=0$ and $\mu=0$, we find 
\be
m_q ^{\text{Ref.}}(T=0,\mu=0)=434\,\text{MeV} \qquad\text{and}\qquad m_q ^{\text{FE}}(T=0,\mu=0)=430\,\text{MeV}
\ee
for $\alpha_{\text{Ref.}}$ and $\alpha_{\text{FE}}$, respectively. For  $T=0$ and $\mu=0$, 
the corresponding scale $k_{\text{cr}}$ at 
which the RG flow enters the regime with broken chiral symmetry is given by
\be
k_{\text{cr}} ^{\text{Ref.}}(T=0,\mu=0)=470\,\text{MeV}\qquad\text{and}\qquad k_{\text{cr}} ^{\text{FE}}(T=0,\mu=0)=330\,\text{MeV}.
\ee
For the phase transition temperature, we find
\be
T_c ^{\text{Ref.}}(\mu=0)=110\,\text{MeV}\qquad\text{and}\qquad T_c ^{\text{FE}}(\mu=0)=76\,\text{MeV}\label{eq:resTc}
\ee
for $\alpha_{\text{Ref.}}$ and $\alpha_{\text{FE}}$, respectively. The difference in the phase transition temperature is
mostly due to the differences in the running of the gauge coupling in the mid-momentum regime~($0.5\,\text{GeV}\lesssim p \lesssim 1.5\,\text{GeV}$). 
Note that the RG flow at finite temperature is less sensitive to the significant difference between the IR fixed point values of the couplings $\alpha_{\text{Ref.}}$ 
and $\alpha_{\text{FE}}$ at vanishing temperature\footnote{This can be seen from looking at the ratios $k_{\text{cr}} ^{\text{FE}}(T\!=\!0,\mu\!=\!0)/k_{\text{cr}} ^{\text{Ref.}}(T\!=\!0,\mu\!=\!0)\approx 0.70$ and $T_c ^{\text{FE}}(\mu\!=\!0)/T_c ^{\text{Ref.}}(\mu\!=\!0)\approx 0.69$. The almost perfect agreement of the two ratios indicates that the
finite-temperature flows are less sensitive to the details of the running of the gauge coupling in the deep IR.}. 
This is because the RG flow of the coupling for $k\lesssim 2\pi T$ is governed by the fixed point of the underlying 3$d$ Yang-Mills theory.
Moreover, chiral symmetry breaking is triggered in the mid-momentum regime rather than in the deep IR regime. 
Therefore we would like to add that we expect in general that our results do not strongly 
depend on the behavior of the coupling in the deep IR, independent of the fact that the IR running of the strong coupling at finite temperature is governed 
by the underlying 3$d$ Yang-Mills theory. To be more specific, the gluons strongly influence the 
matter sector in the mid-momentum regime via gluon-induced four-fermion interactions. As a consequence, the gauge degrees of freedom drive the quark sector 
to criticality depending on the actual temperature, see Subsec.~\ref{SubSec:FixedPoint} for details. Below the chiral symmetry breaking scale 
$k_{\text{cr}}$, the quarks acquire a finite mass. In the deep IR $(k\sim p \lesssim 200\,\text{MeV})$, where currently debated differences between a 
scaling and a decoupling solution in the gauge sector become significant~\cite{Fischer:2008uz}, the quarks are decoupled from the RG flow due to 
their finite mass. Therefore the influence of the gauge-sector on the (chiral) order-parameter potential (i. e. matter sector) in the deep IR is suppressed. 
Thus, chiral symmetry breaking is mostly sensitive to the running of the coupling in the mid-momentum regime rather than its running in the deep IR. 
In this respect, the findings in the present paper are in accordance with the findings in Ref.~\cite{Braun:2007bx} where 
the (de-)confinement phase transition in pure SU(2)- and SU(3)-Yang-Mills theories has been studied.

As discussed in Sec.~\ref{Subsec:RGFlowFourFermi}, we have also checked that the results are only sensitive to our ansatz for the momentum 
dependence of the four-fermion interaction at the percent level. Apart from these errors, systematic errors enter our calculations mainly from omitting instanton 
effects and deconfinement dynamics as described by the Polyakov-Loop. We expect that the transition temperature becomes larger when we include
these effects. Thus the result for $T_c$ given in Eq.~\eqref{eq:resTc} should be considered as a lower bound for the transition temperature. 
We address these issues further after the discussion of our results for the chiral phase boundary.

\begin{table}[t]
\begin{tabular}{l  l  l  c c l}
\hline\hline
Author(s) &  Ref. \hspace*{1.0cm}& Method\hspace*{3.5cm} & $\quad am_c \quad$ & $\quad N_f\quad$ &$\kappa_{\mu}$ \\
 \hline
J. Braun & this work & QCD RG flows & --- & $1$ & $0.97\,..\,1.28$\\
J. Braun & this work & QCD RG flows: $\alpha_s$ vs. $\alpha_{cr}$ & --- & $1$ & $0.40\,..\,0.47$\\
\hline
P. de Forcrand et al.  & \cite{deForcrand:2002ci} & Lattice QCD: imag. $\mu$ & 0.032 & $2$ & $ 0.500(54)$\\
\hline
F. Karsch et al. & \cite{Karsch:2003va} & Lattice QCD: Taylor + Rew. & 0.005 & $3$ & $1.13(45)$\\
P. de Forcrand et al. & \cite{deForcrand:2006pv}  & Lattice QCD: imag. $\mu$ & 0.026 & $3$ & $ 0.667(6)$\\
 \hline\hline
\end{tabular}
\caption{\label{tab:t2_results} Comparison of the results for the curvature of the QCD phase boundary for $N_c=3$ as obtained from Lattice QCD
and QCD RG flows. The table is not exhaustive and we concentrate here on results for degenerate quark flavors with (current) mass $am_c$, where $a$ 
denotes the lattice spacing. The RG calculations have been performed in the chiral limit. Apart from the current quark mass, the Lattice QCD simulations 
differ in the Lattice volumes and spacings used as well as in the implementation of the fermions.}
\end{table}

In order to determine the curvature of the chiral phase boundary of 1-flavor QCD at small quark chemical potential, we compute $T_c(\mu)/T_c(0)$ 
for $0\leq \mu/T_c(0) \lesssim 0.7$.  We then extract the curvature of the phase boundary at vanishing chemical potential from these results. At 
small chemical potential the phase boundary can be expanded in powers of $\mu ^2$, which yields
\be
\frac{T_c(\mu)}{T_c(0)}=1- \kappa_{\mu}(N_f,N_c,m_c) \left(\frac{\mu}{\pi T_c(0)}\right)^2 + \dots\,.
\ee
The coefficient $\kappa_{\mu}$ depends on the number of quark flavors $N_f$, the number of colors $N_c$ 
and the current quark mass. These dependences can be understood qualitatively by looking at the 
underlying mechanisms for chiral symmetry breaking as we discussed them in Sec.~\ref{SubSec:FixedPoint}.
The phase transition temperature and the curvature itself are sensitive to the magnitude of the 
quark condensate at $T=0$ and therefore also to the magnitude of the constituent quark mass at $T=0$. 
The larger the constituent quark mass is for a given chemical potential $\mu$, the smaller is the impact of 
$\mu$ on the phase transition temperature. Therefore $\kappa_{\mu}$ becomes smaller with increasing (current) quark
mass $m_c$. 
\begin{figure}[t]
\includegraphics[scale=1,angle=0]{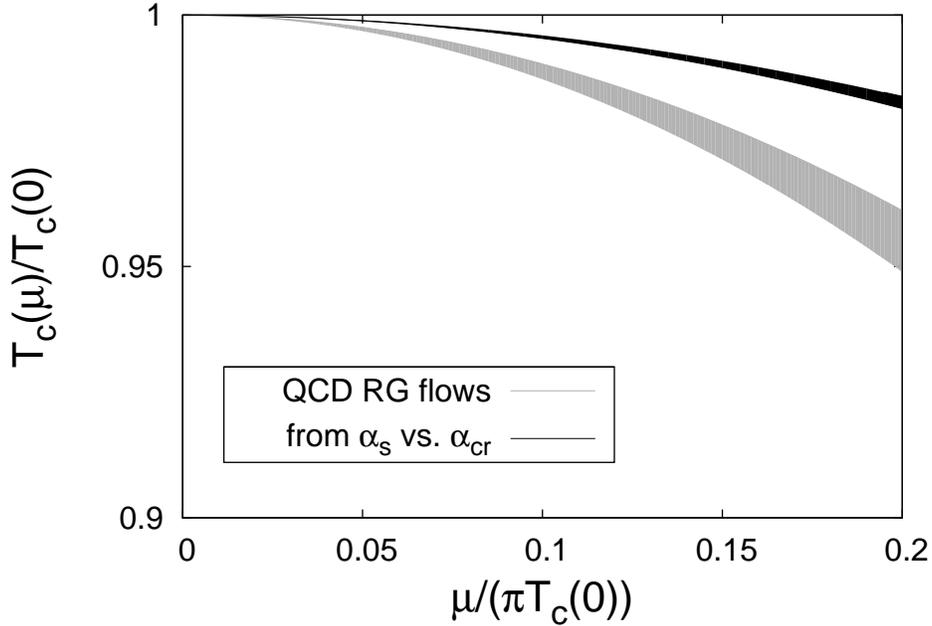}
\caption{Phase boundary of 1-flavor QCD in terms of the dimensionless quantities $T_{cr}(\mu)/T_{cr}(0)$ and $\mu^2/(\pi T_{cr}(0))^2$. The light gray 
band represents the results from the RG flows including the effects of gauge degrees of freedom. The black band indicates the result from an estimate of the
phase boundary from a study of the running of the strong coupling $\alpha_s$ versus $\alpha_{cr}$ along the lines of the study in 
Ref.~\cite{Braun:2005uj,Braun:2006jd}. The width of the bands gives the theoretical error due to uncertainties in the gauge sector. 
\label{Fig:t2_results}}
\end{figure}

The scale for the dynamically generated quark mass is essentially set by the scale $k_{\text{cr}}$ at which the gauge coupling exceeds its critical value. 
The scale $k_{\text{cr}}$ is related to the scale $\Lambda_{\text{QCD}}$ where the gauge coupling strongly increases, thus 
$k_{\text{cr}}\sim\Lambda_{\text{QCD}}$. The flavor- and color-dependence of $\Lambda_{\text{QCD}}$ can be estimated from the position of the Landau 
pole of the perturbative one-loop running of the coupling\footnote{We neglect terms arising from the chemical potential, since $\Lambda_{\text{QCD}}\gg\mu$ for 
the values of the chemical potential we have in mind.} \cite{Braun:2006jd}:
\be
\Lambda_{\text{QCD}}\sim M_Z \,\text{e}^{-\frac{1}{4\pi b_0\alpha(M_Z)}}\approx M_Z\,\text{e}^{-\frac{6\pi}{11 N_c \alpha(M_Z)}}\left(1- \epsilon N_f 
+ {\mathcal O}((\epsilon N_f)^2)  \right)
\ee
where
\be
b_0=\frac{1}{8\pi^2}\left(\frac{11}{3}N_c-\frac{2}{3}N_f\right)\qquad\text{and}\qquad \epsilon= \frac{12\pi}{121 N_c ^2 \alpha(M_Z)}\,.
\ee
We observe that $\Lambda_{\text{QCD}}$ decreases linearly with $N_f$. Therefore the constituent quark mass becomes smaller with increasing $N_f$ 
and $\kappa_{\mu}$ bigger. For fixed $N_f$ but increasing $N_c$, we find that $\Lambda_{\text{QCD}}$ increases because $\ln \frac{\Lambda_{\text{QCD}}}{M_Z}\sim\frac{1}{N_c}$. 
Thus the quark mass increases with $N_c$ and therefore $\kappa_{\mu}$ decreases with increasing $N_c$. A quantitative study of the $N_c$ dependence of the 
phase boundary is work in progress~\cite{Braun:NcDep}. We stress that these simple considerations of the dependences of the curvature in terms of the running 
coupling are in agreement with large $N_c$-considerations of the QCD phase boundary. In Ref.~\cite{Toublan:2005rq}, it has been shown that $\kappa_{\mu}\sim N_f/N_c$ 
in leading order of an expansion in powers of $1/N_c$.

In order to compare our results for the curvature of the phase boundary to results from lattice QCD simulations, we present our result for the phase boundary in 
terms of the curvature $\kappa_{\mu}$. In the first line of Tab.~\ref{tab:t2_results} and Fig.~\ref{Fig:t2_results}, we present our results for the phase boundary for 
QCD with one quark flavor together with results for two- and three-flavor QCD as obtained from lattice QCD simulations. We obtain
\be
\kappa_{\mu}^{\text{Ref.}}=0.97 \qquad\text{and}\qquad \kappa_{\mu}^{\text{FE}}=1.28\label{eq:KappaResults}
\ee
for $\alpha_{\text{Ref.}}$ and $\alpha_{\text{FE}}$, respectively. From this, we conclude that the curvature depends significantly on the gauge-field dynamics whereas 
it is expected that the critical dynamics at the phase transition can be described with a simple $O(2)$ model due to universality.

We list different lattice results for $\kappa_{\mu}$ in Tab.~\ref{tab:t2_results}. From this table, we read off that the lattice result for $\kappa_{\mu}$  for three degenerate quark 
flavors obtained with a Taylor expansion of the path integral is approximately twice as large as the result obtained with imaginary chemical potential. 
From our discussion above, we expect that $\kappa _{\mu}$ decreases roughly linearly with decreasing $N_f$. 
Lattice QCD simulations with imaginary chemical potential are in agreement with this expectation. 
We compare our result for $\kappa_{\mu}(N_f\!=\!1)$ with a naive linear extrapolation of 
$\kappa_{\mu}(N_f\!=\!2)$ and $\kappa_{\mu}(N_f\!=\!3)$ obtained from Lattice simulations with imaginary quark chemical 
potential\footnote{Suitable results for $\kappa_{\mu}$ obtained with sufficiently small quark masses for $N_f=2$ are presently not available 
from Lattice QCD simulations in which a Taylor expansion combined with reweighting techniques has been applied.}
to $N_f=1$. We then find that our result is roughly twice as large as the extrapolated value from these Lattice simulations. 

Let us finally discuss the systematic errors in our results arising from neglected operators associated with instantons and quark confinement. Although our truncation 
allows us to study chiral symmetry breaking with its underlying mechanisms in terms of quarks and gluons, it does not yet allow us to study the deconfinement 
phase transition. Within the RG framework, the deconfinement phase transition in pure Yang-Mills theory with $N_c=2$~\cite{Braun:2007bx,Marhauser2008}, and 
with $N_c=3$~\cite{Braun:2007bx}, has recently been successfully studied. The next step would be to couple the RG flow of the Polyakov to the 
present approach. We can estimate the effects of the inclusion of the Polyakov-loop by considering Landau-DeWitt gauge, $D_{\mu}(\bar{A})(\mathcal{A}-\bar{A})=0$, 
where we identify the background field with the Polyakov-loop field, $\bar{A}=\langle A_0\rangle$. The quark propagator effectively acquires an additional mass term 
since the vacuum expectation value of the zero-component  $\langle A_0\rangle$ of the gauge-field shows up in the Matsubara frequencies of the quark
fields as follows\footnote{We have $\langle A_0\rangle =0$ only for $T\to\infty$.}~\cite{Meisinger:1995ih}:
\be
(\nu_n +\I\mu + \bar{g}\langle A_0\rangle)^2\,.
\ee
As we have discussed above, a larger quark mass translates into a flatter curvature (smaller $\kappa _{\mu}$) and a higher (chiral) phase transition temperature, 
so this would be the likely effect of the Polyakov loop as well. An inclusion of the Polyakov-Loop dynamics in our RG study would eventually allow for a dynamical
study of deconfinement and chiral phase transition and their interplay at the same time.

Instanton effects associated with the $U_A(1)$ anomaly are certainly important for a more quantitative prediction of the curvature of 1-flavor QCD phase boundary since 
they directly influence the quark propagators. Our results for the phase boundary have been obtained from an ansatz for the effective action which has a global $U_A(1)$ 
symmetry. This symmetry is broken in QCD by the presence of gauge-field configurations with non-trivial topology. Instantons are such gauge-field configurations. In the 
context of instantons, the $U_A(1)$ symmetry is broken by their induction of masslike fermion interactions in 1-flavor QCD\footnote{The treatment of instantons within the 
functional RG framework has been discussed in detail in Refs.~\cite{Pawlowski:1996ch,Gies:2002hq}.}~\cite{tHooft:1976fv,Shifman:1979uw,Shuryak:1981ff,Schafer:1996wv}:
\be 
\Gamma _{\mathcal I}=\int d^4 x\, m_{\mathcal I} (\bar{\psi}_R\psi_L - \bar{\psi}_L\psi_R)=\int d^4 x\, m_{\mathcal I} \left(\bar{\psi}\gamma_5\psi\right)\,.
\ee
The associated mass parameter $m_{\mathcal I}$ is exponentially suppressed for small gauge coupling $\alpha_s$, 
$m_{\mathcal I}/\Lambda_{\text{QCD}} \sim \text{e}^{-2\pi/\alpha_s}$, but becomes significantly large near the chiral transition scale, see e.g.~Ref.~\cite{Gies:2002hq}. 
Due to the presence of such an instanton mediated interaction, the pions acquire a mass in the deep IR as well. Therefore the instanton-mediated interactions influence
directly the curvature of the phase boundary and the phase transition temperature. Since the instanton-mediated interactions act like an explicit mass term for the quark fields, 
we expect the curvature to become flatter, the phase transition temperature to become higher and the second order phase transition to turn into a crossover. 

Although we do not study the effects of a broken global $U_A(1)$ symmetry and an inclusion of the Polyakov-Loop dynamics explicitly, we can give an
estimate how they affect the phase boundary by using Eq.~\eqref{eq:TcMuEstimate} to determine the phase boundary.
Estimating the phase boundary from Eq.~\eqref{eq:TcMuEstimate} means to effectively stop the RG flow at the scale $k_{\text{cr}}$ at which the strong coupling exceeds its 
critical value. We find $k_{\text{cr}}\approx 1.5\,\text{GeV}$ for $\alpha_{\text{Ref.}}$ and $k_{\text{cr}}\approx 0.9\,\text{GeV}$ for $\alpha_{\text{FE}}$. Our results 
for the curvature $\kappa_{\mu}$ are given in Tab.~\ref{tab:t2_results}. We find that the estimated values for $\kappa_{\mu}$ are 
\be
\kappa_{\mu}^{\text{Ref.}}=0.47 \qquad\text{and}\qquad \kappa_{\mu}^{\text{FE}}=0.40\,,\label{eq:KappaResults2}
\ee
which are smaller by roughly a factor of two compared to those values obtained from a solution of the full set of flow equations, see Eq.~\eqref{eq:KappaResults} 
and Tab.~\ref{tab:t2_results}. This is due to the fact that the massless Goldstone modes dominating the IR physics are effectively cut off when we estimate the phase 
boundary from Eq.~\eqref{eq:TcMuEstimate}. Instanton-mediated fermionic interactions effectively introduce  an IR cutoff for both the fermions and the bosons 
in the IR which is roughly of the order of $1$ GeV in case of 1-flavor QCD~\cite{Gies:2002hq}. Therefore our estimate of the curvature of the phase boundary from 
Eq.~\eqref{eq:TcMuEstimate} is less contaminated by these topological aspects of QCD and serve as an estimate for a lower bound for the curvature $\kappa_{\mu}$. A 
quantitative analysis of the influence of topological effects on the QCD phase boundary is postponed to future work~\cite{BraunPawlowski:Instanton}.
Moreover we observe that the uncertainty for the curvature in Eq.~\eqref{eq:KappaResults2}, obtained from our estimate for the critical gauge coupling, is much smaller
than the uncertainty found in the results from a solution of the full set of RG flow equations, see Eq.~\eqref{eq:KappaResults}. This is due to the fact that the
running coupling exceeds its critical value on scales where its running is still close to the perturbative 2-loop running and the differences between 
$\alpha_{\text{Ref.}}$ and $\alpha_{\text{FE}}$ are therefore small. On the other hand the results \eqref{eq:KappaResults} for the curvature from the solution of the full set of RG flow 
equations are sensitive to the non-perturbative running of the gauge coupling in the mid-momentum regime ($p\sim 0.5 - 1\;\text{GeV}$). Therefore the 
uncertainty in the curvature $\kappa_{\mu}$ reflects mostly the uncertainty in our truncation of the gauge sector.
\section{Summary and Conclusions}
\label{Sec:Conclusions}
In this paper we have presented a functional RG approach which allows to study the QCD phase boundary from first principles.
Our work aims to set the stage for future studies in this direction incorporating two and three quark flavors.
As a first application, we have computed the phase boundary of 1-flavor QCD at small chemical potential and found 
a second order phase transition if $U_A(1)$ violating terms are neglected. However, our study in its present form 
does not allow us to detect a critical 
endpoint in the phase diagram of 1-flavor QCD since it relies on a low-order expansion in $n$-point functions in the 
scalar sector with $n\leq 4$.  Therefore the reliability (i. e. the radius of convergence) of a Taylor expansion of the 
chiral phase boundary in powers of the quark chemical potential cannot be checked. 
With respect to the nature of the phase transition, however, we expect our truncation to be reliable for small chemical potentials 
(at least for $\mu=0$). In this regime the nature of the finite-temperature phase transition is dominated by the underlying $O(2)$ symmetry 
while quark effects are subleading\footnote{Even though gluon-induced quark interactions drive the system towards the phase transition,
quark effects can be suppressed at the phase transition.}. 
An improvement of our truncation with respect to an inclusion of higher $n$-point functions in the scalar sector of our truncation, 
which allows us to search for a critical endpoint, is deferred to future work.

Apart from a numerical study of the phase boundary, we have discussed the underlying mechanisms of chiral symmetry breaking
in terms of quark-gluon dynamics and how these mechanisms relate to the  dependence of the curvature 
on the number of quark flavors $N_f$, the number of colors $N_c$ and the current quark mass $m_c$. In particular, we have argued 
that the curvature is linearly dependent on $N_f/N_c$.

Although the present approach contains already all ingredients that are necessary for a study of the QCD phase diagram from first principles,
our numerical predictions for the phase boundary at small chemical potentials suffer from the underlying approximations.
As we have already discussed above, our 
present low-order expansion of the scalar sector in terms of $n$-point functions does not allow us to detect a first-order phase transition. 
Moreover, we have further theoretical errors entering our study due to our truncation of the gauge sector. In order
to "measure" the uncertainties arising from this sector we have used the running gauge coupling from a functional RG study employing 
the background-field method~\cite{Braun:2005uj,Braun:2006jd} and from lattice QCD~\cite{Sternbeck:2006cg} as an input. 
Thereby we have exploited the fact that the coupling of the gauge sector and the matter sector is dominantly given by the running 
gauge coupling, the wave-function renormalizations of the gluons as well as the quark mass. We found that the curvature of the 
phase boundaery is about $30\%$ smaller for the lattice coupling than for the background-field coupling. This suggests that 
the curvature of the phase boundary is indeed sensitive to the underlying gauge-field dynamics beyond large $N_c$ which is an important 
result for presently used (P)NJL-type models. With respect to the current debate on scaling versus decoupling scenario in the gauge 
sector, see e. g. Ref.~\cite{Fischer:2008uz},
our analysis suggests that the behavior of the propagators in the deep IR does not strongly influence the shape of the phase boundary. This
observation is accordance with a study of the deconfinement phase transition in pure Yang-Mills theory~\cite{Braun:2007bx}. 

Apart from our analysis of the role of the gauge-field dynamics essentially stemming from the gluonic two-point function, we have 
an uncertainty originating from the fact that we have expanded the gauge-sector about a vanishing $\langle A_0\rangle$ 
instead of taking its temperature and scale-dependence into account~\cite{Braun:2007bx}. Concerning $U_A(1)$-violating terms,
we have provided an estimated of their influence on the phase boundary. We have argued that we expect 
the curvature of the QCD phase diagram at small chemical potentials to become flatter by a factor of two when these two missing pieces are 
included. This lead us to an estimate for a lower bound for the curvature of the phase boundary. We add that this estimate 
for the curvature compares nicely with a linear extrapolation to $N_f=1$ of lattice QCD studies for $N_f=2,3,4$ with imaginary 
chemical potential, see Refs.~\cite{deForcrand:2002ci,D'Elia:2002gd,deForcrand:2003hx,deForcrand:2006pv}. Finally we would like to 
mention that instanton effects as well as an expansion of the gauge sector about a scale and temperature-dependent $\langle A_0\rangle$ 
can be included dynamically in our study by combining it with the findings in Refs.~\cite{Pawlowski:1996ch} and~\cite{Braun:2007bx}, respectively.

We conclude that our approach allows us to compute the phase boundary unambiguously in the sense that the scale for 
all of our results is set by a single input parameter, namely the value of the strong coupling $\alpha_s$ at e. g. the 
Z-boson mass scale. This is a great advantage compared to studies of the QCD phase diagram in terms of (P)NJL-type
models, in which the results for the curvature and the location of a (possible) critical endpoint in the phase diagram 
strongly depend on the choice of the UV cutoff and a set of input parameters. Therefore we think that the present approach is very 
promising since it allows to bridge the gap between quarks and gluons on the one hand and hadronic degrees of freedom on the other 
and opens up the possibility of studies of the QCD phase diagram with two and three quark flavors which depend on only a single physical 
input parameter.  

\begin{acknowledgments}
The author is deeply indebted to Holger Gies, Bertram Klein and Jan Martin Pawlowski for many helpful and enlightening discussions and useful
comments on the manuscript. The author is also grateful to Ian Allison and Philippe de Forcrand for very useful discussions. 
This work was supported by the Natural Sciences and Engineering Research Council 
of Canada~(NSERC). TRIUMF receives federal funding via a contribution agreement 
through the National Research Council of Canada.

\end{acknowledgments}
\appendix
\section{Definition of the Propagators}\label{app:propdef}
We define the boson propagator as follows:
\be
P_{B}(p_0,\{p_i\})=\frac{1}{Z_B ^{\parallel} (p_0,\{p_i\}) p_0 ^2  + Z_B ^{\perp} (p_0,\{p_i\}) \vec{p}^2 (1+r_B) +  M _{B}^2}\,.
\ee
For the fermions, it is convenient to define the modified four-momenta:
\be
\slash\!\!\!\tilde{p}^{(\mp)}&=&\mp \gamma_{\mu}\cdot\left(\frac{Z^{\parallel}_{\psi}(\mp p_0,\{\mp p_i\})}{Z^{\perp}_{\psi}(\mp p_0,\{\mp p_i\})}\frac{\mp p_0 + \I \mu}{1+r_{\psi}}, 
\mp p_i\right)
\equiv \mp\gamma_{\mu}\cdot \tilde{p}_{\mu}^{(\pm)}\,.
\ee
The fermion propagator can then be written as
\be
P_{\psi}^{(+)}(M_{\psi})&=&{\mathcal P}^{(+)}(M_{\psi})\left[\slash\!\!\!\tilde{p}^{(+)} - (1+r_{\psi})^{-1} M_{\psi} \right]\,,
\label{eq:Psi1}\\
P_{\psi} ^{(-)}(M_{\psi})&=&{\mathcal P}^{(-)}(M_{\psi})\left[\slash\!\!\!\tilde{p} ^{(-)} + (1+r_{\psi})^{-1} M_{\psi}\right]
\label{eq:Psi2}
\ee
with
\be
&&{\mathcal P}^{(\mp)}(M_{\psi})\nn\\
&&\qquad\quad=\frac{-Z^{\perp}_{\psi}(\mp p_0,\{\mp p_i\})\,(1+r_{\psi})}
{(Z^{\parallel}_{\psi}(\mp p_0,\{\mp p_i\})\,(\mp p_0 + \I\mu))^2  + (Z^{\perp}_{\psi}(\mp p_0,\{\mp p_i\})\,\vec{p} (1+r_{\psi}))^2 + M_{\psi} ^2}.
\ee
We should keep in mind that $M_{\psi}$ is in general a complicated function depending on the background fields, the Yukawa coupling, the
gauge coupling as well as the four-momentum\footnote{Note that the couplings depend in general on the momenta as well.}:
\be
M_{\psi}\equiv M_{\psi}(p_0,\{p_i\},\{\Phi_i\},h,g,\{{\bar A}_{\mu}\})\,,
\ee
where $\Phi$ represents a two-dimensional vector of real scalar fields. 

Finally, the gauge-field propagator is given by
\be
P_{A}(p_0,\{p_i\})&=&\frac{1}{Z_F ^{\parallel} (p_0,\{p_i\}) p_0 ^2  + Z_F ^{\perp} (p_0,\{p_i\}) \vec{p}^2 (1+r_B)}\left(\delta_{\mu\nu}- \frac{p_{\mu}p_{\nu}}{p^2}\right)\nn\\
&&\qquad\qquad + \frac{\xi}{p_0 ^2  + \vec{p}^2 (1+r_B)}\left( \frac{p_{\mu}p_{\nu}}{p^2}\right)\,,
\ee
where $\xi$ is the gauge-fixing parameter.
\section{Threshold functions}\label{sec:thresholdfcts}
The regulator dependence of the flow equations is controlled by (dimensionless) threshold functions which arise from Feynman graphs, incorporating 
fermionic and/or bosonic Þelds. Let us first introduce the so-called dimensionless regulator-shape function $r_B(x)$ and $r_{\psi}(x)$ for 
bosonic and fermionic fields. These functions  are implicitly defined by the regulator function $R_i$ as follows:
\be
R_B (p_0,\{ p_i\}) = Z^{\perp}_{B,k} (p_0,\{ p_i\}) \vec{p}^{\,2} r_B (\vec{p}^{\,2}/k^2)
\ee
for the bosonic fields and
\be
R_{\psi} (p_0,\{ p_i\}) = Z^{\perp}_{\psi,k} (p_0,\{ p_i\}) \slash\!\!\!\vec{p}\,r_{\psi} (\vec{p}^{\,2}/k^2)
\ee
for the fermionic fields. For the gauge fields, we choose
\be
R_A (p_0,\{ p_i\})=Z^{\perp}_{F,k} (p_0,\{ p_i\})\,\vec{p}^{\,2} r_B (\vec{p}^{\,2}/k^2)\left(\delta_{\mu\nu}- \frac{p_{\mu}p_{\nu}}{p^2}\right)
+\frac{1}{\xi}\,\vec{p}^{\,2} r_B (\vec{p}^{\,2}/k^2)\left( \frac{p_{\mu}p_{\nu}}{p^2}\right)\,.
\ee
In this work, we employ a 3$d$ optimized regulator-shape function \cite{Braun:2003ii,Litim:2006ag,Blaizot:2006rj}:
\be
r_B (x)=\left(\frac{1}{x} - 1\right)\Theta (1-x)\quad\text{and}\quad r_{\psi}=\left(\frac{1}{\sqrt{x}}-1\right)\Theta (1-x)\,.\label{eq:optshapefct}
\ee
In the following, we use these regulator shape functions whenever we evaluate the integrals and sums in our general definitions of the threshold functions.

In order to define the threshold functions, it is convenient to define dimensionless propagators for the scalars (B), gluons (A) and fermions ($\psi$), respectively:
\be
\tilde{G} _{B,A} (x_0,\omega)=\frac{1}{z_{B,A} x_0 + x(1+r_B) + \omega}
\ee 
and
\be
\tilde{G} _{\psi} (x_0,\omega)=\frac{1}{z_{\psi} ^2 x_0 + x(1+r_{\psi})^2 + \omega}\,,
\ee
where $z_{B}=Z^{\parallel}_{B}/Z^{\perp}_{B}$, $z_{A}=Z^{\parallel}_{F}/Z^{\perp}_{F}$ and $z_{\psi}=Z^{\parallel}_{\psi}/Z^{\perp}_{\psi}$ give the
ratio of the wave-function renormalization in the direction parallel and perpendicular to the heat-bath.

First, we define the threshold functions which encounter in the flow equations for the effective potential. For the bosonic loops, we find
\be
l_0 ^{(B),(d)} (\tilde{t},\omega;\eta_B)&=&\frac{\tilde{t}}{2}\sum_{n=-\infty}^{\infty}\int _0 ^{\infty} dx\, x^{\frac{d-1}{2}} 
(\partial _t r_B - \eta_B r_B)\,\tilde{G} _{B} (\tilde{\omega}_n^2,\omega)\nn\\
&=&\frac{2}{d-1}\frac{1}{\sqrt{1+\omega}}\left(1-\frac{\eta_B}{d+1} \right)
\left(\frac{1}{2} + \bar{n}_B(\tilde{t},\omega) \right)\label{eq:l0_BosLoop}
\ee
where $\tilde{t}=T/k$ denotes the dimensionless temperature and $\tilde{\omega}=2\pi n\tilde{t}$ denotes the dimensionless bosonic Matsubara frequencies. 
The function $n_B$ represents the Bose-Einstein distribution function
\be
\bar{n}_B(\tilde{t},\omega)=\frac{1}{\E ^{\sqrt{1+\omega}/\tilde{t}} -1}.
\ee
Bosonic threshold functions of order $n$ are then derived from Eq.~\eqref{eq:l0_BosLoop} by induction:
\be
\frac{\partial}{\partial \omega} l_n ^{(B),(d)} (\tilde{t},\omega;\eta_B) = -(n + \delta_{n,0})\, l_{n+1} ^{(B),(d)} (\tilde{t},\omega;\eta_B)\,.
\ee
For the fermion loops contributing to the flow equations of the effective potential, we find
\be
l_0 ^{(F),(d)} (\tilde{t},\omega,\tilde{\mu};\eta_{\psi})&=&\tilde{t}\sum_{n=-\infty}^{\infty}\int _0 ^{\infty} dx\, x^{\frac{d-1}{2}} 
(\partial _t r_{\psi} - \eta_{\psi} r_{\psi})(1+r_{\psi})\,\tilde{G} _{\psi} ((\tilde{\nu}_n+\I \tilde{\mu})^2,\omega)\nn\\
&=&\frac{1}{d-1}\frac{1}{\sqrt{1+\omega}}\left(1-\frac{\eta_{\psi}}{d} \right)
\left(1 - \bar{n}_{\psi} (\tilde{t},\tilde{\mu},\omega) - \bar{n}_{\psi} (\tilde{t},-\tilde{\mu},\omega) \right)\label{eq:l0_FerLoop}.
\ee
Here, we have introduced the dimensionless fermionic Matsubara frequencies $\tilde{\nu}_n=(2n+1)\pi\tilde{t}$, the dimensionless chemical potential $\tilde{\mu}=\mu/k$ 
and the Fermi-Dirac distribution functions~$n_{\psi}$:
\be
\bar{n}_{\psi} (\tilde{t},\tilde{\mu},\omega) = \frac{1}{\E ^{(\sqrt{1+\omega}-\tilde{\mu})/\tilde{t}} + 1}\stackrel{\tilde{t}\to 0}{\longrightarrow} \Theta(\tilde{\mu}-\sqrt{1+\omega}).
\ee
Higher-order fermionic threshold functions are also given by induction:
\be
\frac{\partial}{\partial \omega} l_n ^{(F),(d)} (\tilde{t},\omega,\tilde{\mu};\eta_{\psi}) = -(n + \delta_{n,0})\, l_{n+1} ^{(F),(d)} (\tilde{t},\omega,\tilde{\mu};\eta_{\psi})\,.
\ee

Let us now define the threshold functions which are involved in the computation of the Yukawa coupling. We have
\be
&&L_{1,1}^{(FB),(d)}(\tilde{t},\omega_{\psi},\tilde{\mu},\omega_{B},\tilde{Q_0};\eta_{\psi},\eta_B)\nn\\
&&\quad=-\frac{\tilde{t}}{2}\,\sum_{n=-\infty}^{\infty}
\int _0 ^{\infty} dx\, x^{\frac{d-3}{2}}\tilde{\partial}_t \,\tilde{G} _{\psi} ((\tilde{\nu}_n+\I \tilde{\mu})^2,\omega_{\psi})\tilde{G} _{B} ((\tilde{\nu}_n-\tilde{Q_0})^2,\omega_{B})\,.
\ee
In order to evaluate the integral over $x$ (spatial momenta), we use\footnote{Here, we give only explicit expressions for the formal derivatives obtained with the regulator 
shape functions~\eqref{eq:optshapefct}.}
\be
\tilde{\partial}_t \Big|_{\psi}&=&\left( \frac{1}{x^{1/2}} - \eta_{\psi}\left(\frac{1}{x^{1/2}} - 1\right)\right)\Theta(1-x)\frac{\partial}{\partial r_{\psi}},\\
\tilde{\partial}_t \Big|_{B}&=&\left( \frac{2}{x} - \eta_{B}\left(\frac{1}{x} - 1\right)\right)\Theta(1-x)\frac{\partial}{\partial r_{\psi}},\\
\ee
where the first and the second line tells us how the formal derivative $\tilde{\partial} _t$ acts on fermions and bosons, respectively. The threshold function $L_{1,1}^{(FB),(d)}$ 
reads then
\be
&& L_{1,1}^{(FB),(d)}(\tilde{t},\omega_{\psi},\tilde{\mu},\omega_{B},\tilde{Q_0};\eta_{\psi},\eta_B)\nn\\
&&\quad =\frac{2\tilde{t}}{d-1}\sum_{n=-\infty}^{\infty} \mathcal{G}_{\psi}((\tilde{\nu}_n+\I \tilde{\mu})^2,\omega_{\psi})\mathcal{G}_B ((\tilde{\nu}_n-\tilde{Q_0})^2,\omega_{B})
\Big\{\left(1-\frac{\eta_{\psi}}{d}\right)\mathcal{G}_{\psi}((\tilde{\nu}_n+\I \tilde{\mu})^2,\omega_{\psi})\nn\\
&& \hspace*{7.5cm} +\left(1-\frac{\eta_{B}}{d+1}\right)\mathcal{G}_{B}((\tilde{\nu}_n - \tilde{Q_0})^2,\omega_{\psi})\Big\},\label{eq:DefL11}
\ee
where we have introduced the auxiliary functions
\be
\mathcal{G}_{\psi}(x_0,\omega)=\frac{1}{z_{\psi} x_0 + 1+ \omega},\\
\mathcal{G}_{B}(x_0,\omega)=\frac{1}{z_B x_0 + 1+ \omega}\,.
\ee
In Landau gauge (or any other gauge with gauge-fixing parameter $\xi \neq1$), we encounter additional threshold functions in the flow equations for the Yukawa coupling 
due to the presence of longitudinal gauge bosons:
\be
&&{\mathcal L}_{1,1}^{(FB),(d)}(\tilde{t},\omega_{\psi},\tilde{\mu},\omega_{B},\tilde{Q_0};\eta_{\psi},\eta_B)\nn\\
&&\quad=-\frac{\tilde{t}}{2}\,\sum_{n=-\infty}^{\infty}
\int _0 ^{\infty} dx\, \frac{x^{\frac{d-1}{2}}}{(\tilde{\nu}_n-\tilde{Q_0})^2 +x}\tilde{\partial}_t \,\tilde{G} _{\psi} ((\tilde{\nu}_n+\I \tilde{\mu})^2,\omega_{\psi})\tilde{G} _{B} ((\tilde{\nu}_n-\tilde{Q_0})^2,\omega_{B})\nn\\
&&\quad =\tilde{t}\sum_{n=-\infty}^{\infty} \mathcal{G}_{\psi}((\tilde{\nu}_n+\I \tilde{\mu})^2,\omega_{\psi})\mathcal{G}_B ((\tilde{\nu}_n-\tilde{Q_0})^2,\omega_{B})
\Big\{\Big( H_{(d-1)}^{(1)} ((\tilde{\nu}_n-\tilde{Q_0})^2)\nn\\ 
&&\qquad\qquad - \eta_{\psi}\left( H_{(d-1)}^{(1)} ((\tilde{\nu}_n-\tilde{Q_0})^2)-H_{d}^{(1)} ((\tilde{\nu}_n-\tilde{Q_0})^2) \right)\Big)\mathcal{G}_{\psi}((\tilde{\nu}_n+\I \tilde{\mu})^2,\omega_{\psi})\nn\\
&&\qquad\qquad\qquad +\Big(  H_{(d-1)}^{(1)} ((\tilde{\nu}_n-\tilde{Q_0})^2)
 - \frac{\eta_{B}}{2}\left( H_{(d-1)}^{(1)} ((\tilde{\nu}_n-\tilde{Q_0})^2)\right. \nn\\
 &&\qquad\qquad\qquad\qquad \left. -H_{(d+1)}^{(1)} ((\tilde{\nu}_n-\tilde{Q_0})^2)  \right)
\Big)\mathcal{G}_{B}((\tilde{\nu}_n - \tilde{Q_0})^2,\omega_B)\Big\},\label{eq:DefLandauL11}
\ee
where the auxiliary functions $H_d(z)$ are given by\footnote{The functions $H_d(z)$ are related to the Hypergeometric function $_2 F_1$, see e. g.~\cite{Gradshteyn}.}
\be
H_{d}^{(m)} (z)=\int _0 ^{1} dx \left(\frac{x^{d/2}}{z +x}\right)^m \,.
\ee

Next, we discuss the threshold functions which are involved in the computation of the wave-function renormalizations. We shall start with the threshold functions for the fermionic 
wave-function $Z_{\psi}$ renormalization due to its close relation to those for the Yukawa coupling. Overall, we have three different types of threshold-functions contributing 
to the flow of $Z_{\psi}$. The first one is closely related to those functions found in Refs.~\cite{Jungnickel:1995fp,Gies:2002hq} and reads:
\be
&&{\mathcal M}_{1,2} ^{(FB),(d)} (\tilde{t},\omega_{\psi},\tilde{\mu},\omega_{B},\tilde{Q_0};\eta_{\psi},\eta_B)\nn\\
&&\;=\frac{\tilde{t}}{2}\,\sum_{n=-\infty}^{\infty}
\int _0 ^{\infty} \!dx  x^{\frac{d-1}{2}}\tilde{\partial}_t \,
\Big\{(1+r_{\psi})\tilde{G} _{\psi} ((\tilde{\nu}_n+\I \tilde{\mu})^2,\omega_{\psi})\frac{d}{dx}\tilde{G} _{B} ((\tilde{\nu}_n-\tilde{Q_0})^2,\omega_{B})\Big\}.
\ee
Evaluating the integration over $x$, we find
\be
&&{\mathcal M}_{1,2} ^{(FB),(d)} (\tilde{t},\omega_{\psi},\tilde{\mu},\omega_{B},\tilde{Q_0})\nn\\
&&\qquad\qquad=\left(1-\frac{\eta_B}{d}\right)\tilde{t}\sum_{n=-\infty}^{\infty}
{\mathcal G} _{\psi} ((\tilde{\nu}_n+\I\tilde{\mu})^2,\omega_{\psi}) \left(\mathcal{G}_{B}((\tilde{\nu}_n-\tilde{Q_0})^2,\omega_B)\right)^2 \,.
\label{eq:M12FB}
\ee
Due to our choice of the regulator functions, ${\mathcal M}_{1,2} ^{(FB),(d)}$ is independent of $Z_{\psi}$, even for \mbox{$\partial_t Z_{\psi}\neq 0$}. The second type of threshold 
function that we encounter is given by
\be
&&{\mathcal N}_{1,2} ^{(FB),(d)} (\tilde{t},\omega_{\psi},\tilde{\mu},\omega_{B},\tilde{Q_0};\eta_{\psi},\eta_B)\\
&&\;=\frac{\tilde{t}}{2}\,\sum_{n=-\infty}^{\infty}
\int _0 ^{\infty} \!dx  \frac{x^{\frac{d-1}{2}+1}}{(\tilde{\nu}_n-\tilde{Q_0})^2+x}\tilde{\partial}_t \,
\Big\{(1+r_{\psi})\tilde{G} _{\psi} ((\tilde{\nu}_n+\I \tilde{\mu})^2,\omega_{\psi})\frac{d}{dx}\tilde{G} _{B} ((\tilde{\nu}_n-\tilde{Q_0})^2,\omega_{B})\Big\}.\nn
\ee
As in the case of the threshold functions for the flow of the Yukawa-coupling, this threshold function as well as the next threshold function that we are going to discuss is only 
present in gauges with gauge-fixing parameter $\xi \neq 1$. Due to the similar momentum structure of ${\mathcal N}_{1,2} ^{(FB),(d)}$ and ${\mathcal M}_{1,2} ^{(FB),(d)}$, we 
immediately obtain
\be
&&{\mathcal N}_{1,2} ^{(FB),(d)} (\tilde{t},\omega_{\psi},\tilde{\mu},\omega_{B},\tilde{Q_0};\eta_{\psi},\eta_B)\nn\\
&&\qquad\qquad=\tilde{t}\sum_{n=-\infty}^{\infty}
{\mathcal G} _{\psi} ((\tilde{\nu}_n+\I\tilde{\mu})^2,\omega_{\psi}) \left(\mathcal{G}_{B}((\tilde{\nu}_n-\tilde{Q_0})^2,\omega_B)\right)^2 
\left(\frac{1}{1+(\tilde{\nu}_n-\tilde{Q_0})^2}\right. \nn\\
&&\hspace*{9cm} \left. -\frac{\eta_B}{2}H_d^{(1)} ((\tilde{\nu}_n-\tilde{Q_0})^2)
\right)
\ee
for the given choice of the regulator functions. The last threshold functions present in the RG flows of $Z_{\psi}$ is a generalization of an already know threshold function 
found in Ref.~\cite{Gies:2002hq}:
\be
&&\tilde{{\mathcal N}}_{1,1,m} ^{(FB),(d)} (\tilde{t},\omega_{\psi},\tilde{\mu},\omega_{B},\tilde{Q_0};\eta_{\psi},\eta_B)\\
&&\,=-\frac{\tilde{t}}{2}\,\sum_{n=-\infty}^{\infty}
\int _0 ^{\infty} \!dx  \frac{x^{\frac{d-3}{2}+m}}{((\tilde{\nu}_n-\tilde{Q_0})^2 \!+\! x)^m}\tilde{\partial}_t \,
\Big\{(1+r_{\psi})\tilde{G} _{\psi} ((\tilde{\nu}_n+\I \tilde{\mu})^2,\omega_{\psi})\tilde{G} _{B} ((\tilde{\nu}_n-\tilde{Q_0})^2,\omega_{B})\Big\}.\nn
\ee
The integral over $x$ can be performed straightforwardly, yielding
\be
&&\tilde{{\mathcal N}}_{1,1,m} ^{(FB),(d)} (\tilde{t},\omega_{\psi},\tilde{\mu},\omega_{B},\tilde{Q_0};\eta_{\psi},\eta_B)\nn\\
&&\;=-\frac{\tilde{t}}{2}\sum_{n=-\infty}^{\infty}
{\mathcal G} _{\psi} ((\tilde{\nu}_n+\I\tilde{\mu})^2,\omega_{\psi}) \mathcal{G}_{B}((\tilde{\nu}_n-\tilde{Q_0})^2,\omega_B)
\Bigg\{
H_{(d-4)}^{(m)}((\tilde{\nu}_n\!-\!\tilde{Q_0})^2) \nn\\
&&\;\;\quad -\eta_{\psi}\Big[ H_{(d-4)}^{(m)}((\tilde{\nu}_n\!-\!\tilde{Q_0})^2) - H_{(d-3)}^{(m)}((\tilde{\nu}_n\!-\!\tilde{Q_0})^2)\Big]
-2 {\mathcal G} _{\psi} ((\tilde{\nu}_n\! +\!\I\tilde{\mu})^2,\omega_{\psi}) H_{(d-4)}^{(m)}((\tilde{\nu}_n\!-\!\tilde{Q_0})^2)\nn\\
&&\quad\quad +2\eta_{\psi} {\mathcal G} _{\psi} ((\tilde{\nu}_n\! +\!\I\tilde{\mu})^2,\omega_{\psi}) \Big[
H_{(d-4)}^{(m)}((\tilde{\nu}_n\!-\!\tilde{Q_0})^2) - H_{(d-3)}^{(m)}((\tilde{\nu}_n\!-\!\tilde{Q_0})^2)
\Big]\nn\\
&&\quad\quad\quad -2\mathcal{G}_{B}((\tilde{\nu}_n\!-\!\tilde{Q_0})^2,\omega_B) H_{(d-4)}^{(m)}((\tilde{\nu}_n\!-\!\tilde{Q_0})^2)
+\eta_B \mathcal{G}_{B}((\tilde{\nu}_n\!-\!\tilde{Q_0})^2,\omega_B)\Big[ H_{(d-4)}^{(m)}((\tilde{\nu}_n\!-\!\tilde{Q_0})^2)
\nn\\
&&\quad\quad\quad\quad - H_{(d-2)}^{(m)}((\tilde{\nu}_n\!-\!\tilde{Q_0})^2)\Big]\Bigg\}\,.
\ee

The threshold functions needed for the computation of the scalar anomalous dimensions are given by
\be
{\mathcal M}_{2,2} ^{(B),(d)}(\tilde{t},\omega_1,\omega_2;\eta_{B})
=-\frac{\tilde{t}}{2}\sum_{n=-\infty}^{\infty}\int _0 ^{\infty} \!dx x^{\frac{d-1}{2}}\tilde{\partial}_t
\left( \frac{d}{dx} \tilde{G} _{B} (\tilde{\omega}_n^2,\omega_{1})\right)\!\!\!\left( \frac{d}{dx} \tilde{G} _{B} (\tilde{\omega}_n^2,\omega_{2})\right).
\ee
The integration over $x$ can be performed analytically for the regulator shape functions under consideration 
and we obtain
\be
{\mathcal M}_{2,2} ^{(B),(d)}(\tilde{t},\omega_1,\omega_2;\eta_{B})=\tilde{t}\sum_{n=-\infty}^{\infty}
\left(\mathcal{G}_{B}(\tilde{\omega}_n^2,\omega_1)\right)^2 \left(\mathcal{G}_{B}(\tilde{\omega}_n^2,\omega_2)\right)^2\,.
\label{eq:M22B}
\ee
Note that ${\mathcal M}_{2,2} ^{(B),(d)}$ is independent of $\eta_\phi$ for our choice of the regulator functions as it has already been found in Ref.~\cite{Hofling:2002hj}. 
Two more contributions to the scalar anomalous dimensions come from purely fermionic loops. One of those contributions is proportional to the square of the vacuum
expectation value of the scalar field and the corresponding threshold function reads
\be
{\mathcal M}_{2} ^{(F),(d)}(\tilde{t},\omega,\tilde{\mu};\eta_{\psi})
=-\frac{\tilde{t}}{2}\sum_{n=-\infty}^{\infty}\int _0 ^{\infty} \!dx x^{\frac{d-1}{2}}\tilde{\partial}_t
\left( \frac{d}{dx} \tilde{G} _{\psi} ((\tilde{\nu}_n+\I\tilde{\mu})^2,\omega)\right)^2.
\ee
The integration over $x$ can be carried out analytically and we find again that ${\mathcal M}_{2,2} ^{(F),(d)}$ is independent of the fermionic anomalous dimensions for 
the cutoff function under consideration:
\be
{\mathcal M}_{2} ^{(F),(d)}(\tilde{t},\omega,\tilde{\mu};\eta_{\psi})=\tilde{t}\sum_{n=-\infty}^{\infty}
\left(\mathcal{G}_{\psi}((\tilde{\nu}_n+\I \tilde{\mu})^2,\omega)\right)^4\,.
\label{eq:M2F}
\ee
The threshold function of the second contribution consisting solely of fermionic internal lines is given by
\be
{\mathcal M}_{4} ^{(F),(d)}(\tilde{t},\omega,\tilde{\mu};\eta_{\psi})
=-\frac{\tilde{t}}{2}\sum_{n=-\infty}^{\infty}\int _0 ^{\infty} \!dx x^{\frac{d+1}{2}}\tilde{\partial}_t
\left( \frac{d}{dx} (1+r_{\psi})\tilde{G} _{\psi} ((\tilde{\nu}_n+\I\tilde{\mu})^2,\omega)\right)^2.
\ee
Performing the integration over $x$, we find
\be
{\mathcal M}_{4} ^{(F),(d)}(\tilde{t},\omega,\tilde{\mu};\eta_{\psi})
&=&\tilde{t}\sum_{n=-\infty}^{\infty}
\Big\{ \left(\mathcal{G}_{\psi}((\tilde{\nu}_n+\I \tilde{\mu})^2,\omega)\right)^4
+\frac{1-\eta_{\psi}}{d-3} \left(\mathcal{G}_{\psi}((\tilde{\nu}_n+\I \tilde{\mu})^2,\omega)\right)^3 \nn\\
&&\qquad\qquad\qquad - \left(\frac{1-\eta_{\psi}}{2d-6} +\frac{1}{4}\right)\left(\mathcal{G}_{\psi}((\tilde{\nu}_n+\I \tilde{\mu})^2,\omega)\right)^2\Big\}.
\label{eq:M4F}
\ee

Finally, we have to discuss the threshold functions corresponding to 1 PI box diagrams. These threshold functions are needed for a computation of the RG flow of the four-fermion 
interaction. As in the case of the Yukawa-coupling and the fermionic wave-function renormalization, we encounter additional threshold-functions in gauges with
gauge-fixing parameter $\xi \neq 1$. We start, however, with the discussion of the threshold functions that are closely related to those already discussed in 
Refs.~\cite{Meggiolaro:2000kp,Gies:2001nw,Gies:2002hq}:
\be
&& L^{(FB),(d)}_{1,1,n_1,n_2} (\tilde{t},\omega_{\psi},\tilde{\mu}_1,\tilde{\mu}_2,\omega_{B,1},\omega_{B,2};\eta_{\psi},\eta_B)\nn\\
&&\;\;=-\frac{\tilde{t}}{2}\sum _{n=-\infty}^{\infty}\int _0 ^{\infty} dx x^{\frac{d-1}{2}}\tilde{\partial}_t\left\{
\Big[ (1+r_{\psi})\tilde{G} _{\psi} ((\tilde{\nu}_n+\I\tilde{\mu}_1)^2,\omega_{\psi})\Big]\Big[ (1+r_{\psi})\tilde{G} _{\psi} ((\tilde{\nu}_n+\I\tilde{\mu}_2)^2,\omega_{\psi})\Big]\times\right.\nn\\
&&\hspace*{8.0cm}\left.\times\Big[ \tilde{G} _{B} (\tilde{\nu}_n^2,\omega_{B,1})\Big]^{n_1}\Big[ \tilde{G} _{B} (\tilde{\nu}_n^2,\omega_{B,2})\Big]^{n_2}\right\}.
\ee
By performing the integration over $x$, we obtain the expression for the threshold function used in this work:
\be
&& L^{(FB),(d)}_{1,1,n_1,n_2} (\tilde{t},\omega_{\psi},\tilde{\mu}_1,\tilde{\mu}_2,\omega_{B,1},\omega_{B,2};\eta_{\psi},\eta_B)\nn\\
&&=\frac{2\,\tilde{t}}{d-1}\sum_{n=-\infty}^{\infty}
\mathcal{G}_{\psi}((\tilde{\nu}_n+\I \tilde{\mu}_1)^2,\omega_{\psi})\mathcal{G}_{\psi}((\tilde{\nu}_n+\I \tilde{\mu}_2)^2,\omega_{\psi})
\Big[\mathcal{G}_{B}(\tilde{\nu}_n^2,\omega_{B,1})\Big]^{n_1}\Big[\mathcal{G}_{B}(\tilde{\nu}_n^2,\omega_{B,2})\Big]^{n_2}\times\nn\\
&&\;\;\times\Bigg\{
\Big(\mathcal{G}_{\psi}((\tilde{\nu}_n+\I \tilde{\mu}_1)^2,\omega_{\psi}) + \mathcal{G}_{\psi}((\tilde{\nu}_n+\I \tilde{\mu}_2)^2,\omega_{\psi})
+n_1\, \mathcal{G}_{B}(\tilde{\nu}_n^2,\omega_{B,1}) + n_2\, \mathcal{G}_{B}(\tilde{\nu}_n^2,\omega_{B,2}) - 1\Big)\nn\\
&&\qquad\qquad-\frac{\eta_B}{d+1}\Big( n_1\,\mathcal{G}_{B}(\tilde{\nu}_n^2,\omega_{B,1}) + n_2\, \mathcal{G}_{B}(\tilde{\nu}_n^2,\omega_{B,2}) \Big)\nn\\
&&\qquad\qquad\qquad-\frac{\eta_{\psi}}{d}\Big(\mathcal{G}_{\psi}((\tilde{\nu}_n+\I \tilde{\mu}_1)^2,\omega_{\psi}) + \mathcal{G}_{\psi}((\tilde{\nu}_n+\I \tilde{\mu}_2)^2,\omega_{\psi}) -1\Big)
\Bigg\}.
\label{eq:L11n1n2FB}
\ee
Since we are explicitly studying Landau-gauge QCD, we have an additional class of threshold functions given by
\be
&& {\mathcal L}^{(FB),(d)}_{1,1,n} (\tilde{t},\omega_{\psi},\tilde{\mu}_1,\tilde{\mu}_2,\omega_{B};\eta_{\psi},\eta_B)\nn\\
&&\qquad=-\frac{\tilde{t}}{2}\sum _{n=-\infty}^{\infty}\int _0 ^{\infty} dx \frac{x^{\frac{d+1}{2}}}{\tilde{\nu}^2 \!+\! x}\tilde{\partial}_t\left\{
\Big[ (1\!+\! r_{\psi})\tilde{G} _{\psi} ((\tilde{\nu}_n\!+\!\I\tilde{\mu}_1)^2,\omega_{\psi})\Big]\times\right.\nn\\
&&\hspace*{5.25cm}\left.\times\Big[ (1\!+\! r_{\psi})\tilde{G} _{\psi} ((\tilde{\nu}_n\!+\!\I\tilde{\mu}_2)^2,\omega_{\psi})\Big]\Big[ \tilde{G} _{B} (\tilde{\nu}_n^2,\omega_{B})\Big]^{n}\right\}.
\ee
Inserting the regulator shape functions~\eqref{eq:optshapefct}, we find
\be
&&\hspace*{-0.5cm} {\mathcal L}^{(FB),(d)}_{1,1,n} (\tilde{t},\omega_{\psi},\tilde{\mu}_1,\tilde{\mu}_2,\omega_{B};\eta_{\psi},\eta_B)\nn\\
&&\hspace*{-0.25cm} =\tilde{t}\sum_{n=-\infty}^{\infty}
\mathcal{G}_{\psi}((\tilde{\nu}_n\!+\!\I \tilde{\mu}_1)^2,\omega_{\psi})\mathcal{G}_{\psi}((\tilde{\nu}_n\!+\!\I \tilde{\mu}_2)^2,\omega_{\psi})
\Big[\mathcal{G}_{B}(\tilde{\nu}_n^2,\omega_{B})\Big]^{n}\times\nn\\
&&\times\Bigg\{
H^{(1)}_{(d-1)}(\tilde{\nu}_n ^2)\Big(\mathcal{G}_{\psi}((\tilde{\nu}_n\!+\!\I \tilde{\mu}_1)^2,\omega_{\psi}) + \mathcal{G}_{\psi}((\tilde{\nu}_n\!+\!\I \tilde{\mu}_2)^2,\omega_{\psi})
+n\, \mathcal{G}_{B}(\tilde{\nu}_n^2,\omega_{B}) -1\Big)\nn\\
&&\;\; -\frac{\eta_B}{2}\left(H^{(1)}_{(d-1)}(\tilde{\nu}_n ^2)\! -\! H^{(1)}_{(d+1)}(\tilde{\nu}_n ^2)\right) n\,\mathcal{G}_{B}(\tilde{\nu}_n^2,\omega_{B}) \nn\\
&&\;\;\quad -\eta_{\psi}\left(H^{(1)}_{(d-1)}(\tilde{\nu}_n ^2)\! -\! H^{(1)}_{d}(\tilde{\nu}_n ^2)\right)\!\!
\Big(\mathcal{G}_{\psi}((\tilde{\nu}_n\!+\!\I \tilde{\mu}_1)^2,\omega_{\psi})\! +\! \mathcal{G}_{\psi}((\tilde{\nu}_n\!+\!\I \tilde{\mu}_2)^2,\omega_{\psi})\! -\! 1\Big)
\!\Bigg\}.
\ee

\bibliography{QCD_phaseboundary}

\end{document}